\documentclass[iop]{emulateapj}

\shorttitle{Central kpc in Seyferts versus Quiescent Galaxies}
\shortauthors{Hicks et al.}

\def \as{\hbox{$^{\prime\prime}$}}

\def \Lsun{L$_{\odot}$}

\def \mic{$\mu$m}
\def \lam{$\lambda$}
\def \sig{$\sigma$}
\def \it{\textit{}}
\def \htwo{H$_{2}$}
\def \htwos{H$_{2}$ 1-0~S(1)}
\def \kms{km s$^{-1}$}
\def \sm{$\sim$}
\def \kb{{\em K}-band}
\def \deg{\,\hbox{$^\circ$}}

\begin{document}

\title{Fueling Active Galactic Nuclei I. How the Global Characteristics of the Central Kiloparsec of Seyferts differ from Quiescent Galaxies$^{*}$}

\author{E. K. S. Hicks\altaffilmark{1}, R. I. Davies\altaffilmark{2}, W. Maciejewski\altaffilmark{3}, E. Emsellem\altaffilmark{4}, M. A. Malkan\altaffilmark{5}, G. Dumas\altaffilmark{6}, F. M{\"u}ller-S{\'a}nchez\altaffilmark{5}, A. Rivers\altaffilmark{1}}

\footnotetext[*]{Based on observations at the ESO Very Large Telescope (083.B-0332).}
\altaffiltext{1}{Department of Astronomy, Box 351580, University of Washington Seattle, WA 98195, USA; ehicks@astro.washington.edu}
\altaffiltext{2}{Max Planck Institut f{\"u}r extraterrestrische Physik, Postfach 1312, 85741 Garching, Germany}
\altaffiltext{3}{Astrophysics Research Institute, Liverpool John Moores University, Twelve Quays House, Egerton Wharf, Birkenhead CH41 1LD, UK}
\altaffiltext{4}{European Southern Observatory, 85748 Garching bei Muenchen, Germany}
\altaffiltext{5}{Astronomy Division, University of California, Los Angeles, CA 90095-1562, USA}
\altaffiltext{6}{Institut de Radio Astronomie Millim{\'e}trique (IRAM), 300, rue de la Piscine, Domaine Universitaire, 38406 Saint Martin dHeres,
France}

\begin{abstract}
In a matched sample of Seyfert and quiescent galaxies we simultaneously probe the stellar and molecular gas kinematics from 1 kpc down to 50 pc with the aim of identifying the dynamical processes dictating black hole accretion rates.  This first paper compares the global characteristics of a sample of ten galaxies.  We find several differences within a radius of 500 pc that are correlated with Actkve Galactic Nucleus (AGN) activity.  The Seyferts have: (1) a more centrally concentrated nuclear stellar surface brightness with lower stellar luminosities beyond a radius of 100 pc, (2) a lower stellar velocity dispersion within a radius of 200 pc, (3) elevated \htwos~luminosity out to a radius of at least 250 pc, and (4) more centrally concentrated \htwo~surface brightness profiles.  These observed differences can be interpreted as evidence for Seyfert galaxies having a dynamically cold (in comparison to the bulge) nuclear structure composed of a significant gas reservoir and a relatively young stellar population.  This structure is undetected (and possibly does not exist) in quiescent galaxies.  The presence of such a nuclear structure in Seyfert galaxies provides evidence for inflow of the surrounding interstellar medium since the nuclear stellar population requires a supply of gas from which to form.  The fueling of a Seyfert AGN is thus associated with the formation of a dynamically cold component of gas and stars on scales of hundreds of parsecs.

\end{abstract}

\keywords{
galaxies: active --- 
galaxies: kinematics and dynamics ---
galaxies: nuclei ---
galaxies: Seyfert ---
infrared: galaxies} 

\section{Introduction}
\label{sec:intro}

We have now come to the understanding that supermassive black holes (BHs) play a pivotal role in galaxy evolution.  This revelation has been driven by the discovery that most, if not all, galaxies harbor a BH, and that the masses of these BHs are related to the global properties of their host galaxy (e.g. \citealt{ferrarese00}, \citealt{gebhardt00}, \citealt{tremaine02}).  This relationship is thought to be a consequence of the coevolution of the BH and its host galaxy, and identifying the origin of this coevolution is an essential component to our understanding of galaxy formation and evolution.  It is widely accepted that BHs accumulate their mass through an Active Galactic Nuclei (AGN) phase in which they accrete material from the host galaxy.  \citet{kauffmann03} find that Seyfert AGN generally reside in host galaxies with a younger stellar population relative to that found in quiescent galaxies, confirming that an abundant fuel supply is available in these galaxies on scales of several kiloparsecs.  However, we have yet to identify the processes responsible for actually driving material down to the radii necessary for accretion onto the BH.  We must therefore first identify and understand these processes before forming a complete picture of the role of BHs in the evolution of galaxies.

Major mergers have been shown to be an efficient mechanism for rapid loss of angular moment that can support the fueling of the most luminous quasar AGN (e.g. \citealt{barnes96}, \citealt{hopkins08}).  However, it is likely that this merger driven inflow is stalled at scales of 1 kpc, at which point it is actually smaller scale processes that are expected to be responsible for directly fueling the central BH (e.g. \citealt{hopkins10}).  In addition, multiwavelength surveys find that it is only the most luminous AGN that are associated with major mergers leaving a significant fraction of AGN out to a redshift of at least z = 3 to be driven by other processes (e.g. Treister et al. 2012).  Recent studies have also shown that at redshifts of z $>$ 1 a significant fraction of AGN reside in disk galaxies, which strongly suggests that secular disk processes rather than major mergers are the principle driver fueling their AGN (e.g. \citealt{schawinski11,schawinski12}, \citealt{kocevski12}).  By redshifts of z $\sim$ 1-1.5 disk galaxies begin to show the characteristics of the local Hubble sequence (\citealt{vandenbergh00}, \citealt{kajisawa01}, \citealt{conselice04}, \citealt{oesch10}) and bars can be traced out to z $\sim$ 1 (\citealt{sheth08}), thus secular processes become a promising alternative source of AGN fueling by these redshifts.  Although secular processes capable of driving the needed quantities of fuel inward have been identified, e.g. large-scale spiral structures and bars, these processes are only thought to operate down to scales of hundreds of parsecs.  This may explain why previous studies of large-scale structures in local galaxies have found little, or no, correlation with AGN activity (e.g. \citealt{shlosman00}, \citealt{laine02}, \citealt{laurikainen02,laurikainen04}, \citealt{hao09}).  It is thus a series of processes that carry material inward (\citealt{shlosman90}, \citealt{wada04}) and it is the smaller scale processes (less than 1 kpc) that are ultimately regulating BH growth in the local universe as well as at high redshifts.  

To constrain what processes are relevant to fueling local AGN, and are likely candidates for fueling at higher redshifts as well, we have observed a sample of matched Seyfert-quiescent galaxy pairs from which we can identify properties unique to Seyferts that are potentially related to fueling of their AGN activity.  A significant impediment to addressing this issue thus far has been the limitation of relying on morphological signatures of these processes, while the critical difference between Seyfert and inactive galaxies is expected to lie in the kinematics of the interstellar medium (ISM).  Comparison studies of the circumnuclear dust structures in the central kpc of galaxies imaged with the Hubble Space Telescope (HST) have found little or no correlation between isophotal twists, nuclear bars, and nuclear spirals detected in Seyferts relative to inactive galaxies (e.g. \citealt{hunt04}, \citealt{Martini03a,Martini03b}).  \citet{simoes07} find no correlation between the presence of circumnuclear dust and AGN in late-type galaxies (the typical hosts of Seyfert AGN), but do find a correlation in early-type galaxies.  However, they conclude that even in early-type galaxies this structure is not a sufficient condition for AGN activity since similar structure is also seen in a significant fraction of comparable quiescent galaxies, and that there must be other critical factors in driving AGN.

Extending beyond morphological signatures of AGN fueling, \citet{dumas07} carried out a 2D kinematic comparative study of a sample of Seyfert and inactive galaxies and report that on scales of 200 pc the kinematics of the rotating disk of ionized gas in Seyfert galaxies deviate from axisymmetric rotation to a greater degree than in inactive galaxies.  They conclude that the link between the AGN and the host galaxy dynamics therefore exists on scales of hundreds to tens of parsecs.  This suggests that these are the dynamical timescales of the fueling mechanism(s) that are relevant to Seyfert activity.  Looking at the cold ISM directly using 0.5-1\as~resolution CO data, the NUGA team have assessed the potential for gravitational torques fueling AGN and find that only 1/3 of a sample of 25 galaxies are driving gas into the central 100 pc (\citealt{haan09}, \citealt{garcia12}).  However, a large fraction of the galaxies in this sample are low luminosity AGN with LINER or transition object classifications, rather than exclusively Seyferts.  It is therefore possible that many of the galaxies in the NUGA sample can be driven by more subtle fueling mechanisms on smaller scales that do not represent the fueling requirements of higher luminosity local and higher redshift AGN.  

To constrain what mechanisms are responsible for fueling of local AGN, and potentially AGN at higher redshifts, we have carried out a matched Seyfert-quiescent galaxy study in which we measure the 2D kinematics of molecular hydrogen that is expected to trace the cooler ISM fueling AGN (e.g. \citealt{hicks09}, \citealt{sb10}, \citealt{riffelsb11}).  In this first of a series of papers, we present a global analysis of the properties of Seyferts compared to quiescent galaxies and identify several differences in the nuclear (within a radius of 250 pc) stellar and molecular gas properties.  Future papers will present a more detailed analysis of the kinematics in the observed galaxies, including modeling to reliably demonstrate and quantify the detected ISM inflow.  Our observed sample, data processing techniques, and confirmation of nuclear activity in the selected galaxies is discussed in $\S$ 2.  In $\S$ 3 the distribution and kinematics of both the nuclear stars and molecular gas are presented.  Results based on our comparative analysis of the global properties in the Seyfert and quiescent galaxy subsamples are discussed in $\S$ 4, and in $\S$ 5 evidence for a lack of bulk \htwo~outflows and subsample differences in the detection of ionized gas emission are considered.  This is followed by an interpretation of the observed subsample differences in $\S$ 6, and conclusions are presented in $\S$ 7.

\section{Sample Selection, Observations, and Data Processing}

\subsection{Matched Seyfert-Quiescent Galaxy Sample}
\label{sec:sample}
In order to identify differences between Seyfert and quiescent galaxies that can provide insight into what mechanisms are fueling the Seyfert AGN the observed sample was selected to reduce any potential biases due to large-scale ($>$\,kpc) host galaxy properties.  The sample is therefore draw from the morphological study of \citet{Martini03a,Martini03b} who matched a sample of 26 AGN and inactive galaxies pairs imaged with HST based on the host galaxy's Hubble type, {\em B}-band luminosity, heliocentric velocity, inclination, and angular size (see Table \ref{tab:galprops} for matching criteria).  From this sample, pairs have been selected in which both galaxies are visible from VLT and have $<$\,200\,pc/$''$ (distance\,$<$\,40\,Mpc), and for which the AGN is reported to be a Seyfert in \citet{Martini03a}.  To ensure that differences in the sub-samples are linked to AGN activity level, weaker levels of nuclear AGN activity (i.e. those galaxies classified as LINERs or transition galaxies in \citealt{Martini03a}) have been avoided to maintain a clear distinction between activity resulting from fueling of a BH and more global activity, such as vigorous star formation.  The observed sample is comprised of 10 galaxies, five Seyfert and five quiescent galaxies, the host galaxy properties for which are given in Table \ref{tab:galprops}.  It is worth noting that such a matched sample is not strictly necessary since, as discussed in $\S$\ref{sec:intro}, no large-scale properties have actually been found to correlate with Seyfert AGN and therefore any differences are not expected to actually bias the samples.  The observed sample is therefore also treated as two subsamples of five active and five quiescent galaxies, in addition to considering a matched pair-wise comparison.

\begin{deluxetable*}{cllcccccl}
\tabletypesize{\small}
\tablecaption{Galaxy Properties \label{tab:galprops}} 
\tablewidth{0pt}
\tablehead{
\colhead{I.D.} &
\colhead{Galaxy} &
\colhead{AGN} &
\colhead{{\em T}} &
\colhead{{\em M}$_{B}$} &
\colhead{{\em v}} &
\colhead{{\em R}$_{25}$} &
\colhead{{\em frac}$_{20}$} &
\colhead{Nuc. Dust} \\
\colhead{(1)} &
\colhead{(2)} &
\colhead{(3)} &
\colhead{(4)} &
\colhead{(5)} &
\colhead{(6)} &
\colhead{(7)} &
\colhead{(8)} &
\colhead{(9)} \\
}
\startdata
1a  & NGC 3227 & Sey 1.5  & 1.0 & -20.4 & 1152 & 0.17 & 0.06 & C \\
1q  & IC 5267  & no       & 0.0 & -20.3 & 1713 & 0.13 & 0.06 & C \\
2a  & NGC 5643 & Sey 2    & 5.0 & -20.9 & 1199 & 0.06 & 0.07 & GD \\
2q  & NGC 4030 & no       & 4.0 & -20.9 & 1460 & 0.14 & 0.16 & TW \\
3a  & NGC 6300 & Sey 2    & 3.0 & -20.6 & 1110 & 0.18 & 0.07 & C \\
3q  & NGC 3368 & no       & 2.0 & -20.3 & \phn897 & 0.16 & 0.04 & CS \\
4a  & NGC 6814 & Sey 1.5  & 4.0 & -20.5 & 1563 & 0.03 & 0.11 & GD \\
4q  & NGC  628 & no       & 5.0 & -20.2 & \phn 656 & 0.04 & 0.03 & N \\
5a  & NGC 7743 & Sey 2    & -1.0 & -19.4 & 1658 & 0.07 & 0.11 & LW \\
5q  & NGC  357 & no       & 0.0 & -19.9 & 2541 & 0.14 & 0.14 & N \\
\hline
\multicolumn{3}{l}{Seyfert Subsample}   & 2.4      & -20.4    & 1336     & 0.10 & 0.05 & \\
  &    &                                & $\pm$2.4 & $\pm$0.6 & $\pm$254 & $\pm$0.07 & $\pm$0.02 & \\
\multicolumn{3}{l}{Quiescent Subsample} & 2.2      & -20.3    & 1453     & 0.12 & 0.10 & \\
  &    &                                & $\pm$2.3 & $\pm$0.4 & $\pm$741 & $\pm$0.05 & $\pm$0.04 & \\

\enddata
\tablecomments{Galaxy I.D.s given in {\em column 1}, and used throughout the paper, are based on the matched pairs as described in $\S$ \ref{sec:sample}, for example the first pair is the ``active" Seyfert galaxy 1a and the ``quiescent" galaxy 1q.  Host galaxy properties in {\em columns 4-8} are as reported in \citet{Martini03a} and the respective matching criteria for galaxy pairs are as follows: Hubble Type $\Delta${\em T} $<$ 1, blue luminosity $\Delta${\em M}$_B$ $<$ 0.5 mag, heliocentric velocity $\Delta${\em v} $<$ 1000 \kms, inclination $\Delta${\em R}$_{25}$ $<$ 0.1 (defined as the mean decimal logarithm of the ratio of the major isophotal diameter to the minor isophotal diameter), and the fraction of the angular diameter D$_{25}$ of the host galaxy within the 20\as~field of view of the {\em HST} maps published in \citealt{Martini03a} $\Delta${\em frac}$_{20}$ $<$ 0.1.  {\em Column 9} gives the nuclear dust structure as classified by \citealt{Martini03a} where the classification codes are GD for grand-design nuclear spiral, TW for tightly wound spiral, LW for loosely wound spiral, CS for chaotic spiral, C for chaotic circumnuclear dust, and N for no circumnuclear dust structure.  Average values for each subsample are given in the bottom rows with the mean in the top row and the standard deviation in the row immediately below.}

\end{deluxetable*} 


\begin{deluxetable*}{cllcccccc}
\tabletypesize{\scriptsize}
\tablecaption{Galaxy Properties \label{tab:galprops2}} 
\tablewidth{0pt}
\tablehead{
\colhead{I.D.} &
\colhead{Galaxy} &
\colhead{AGN} &
\colhead{{\em L}$_{2-10 keV}$} &
\colhead{{\em H}-band (iso.)} &
\colhead{{\em H}-band (total)} &
\multicolumn{2}{c}{0.5$D_{25}$} &
\colhead{B/T} \\
\colhead{} &
\colhead{} &
\colhead{} &
\colhead{(10$^5$ \Lsun)} &
\colhead{(mag)} &
\colhead{(mag)} &
\colhead{(\arcsec)} &
\colhead{(kpc)} & 
\colhead{} \\
\colhead{(1)} &
\colhead{(2)} &
\colhead{(3)} &
\colhead{(4)} &
\colhead{(5)} &
\colhead{(6)} &
\colhead{(7)} &
\colhead{(8)} &
\colhead{(9)} \\
}
\startdata
1a  & NGC 3227 & Sey 1.5  & 4141          & 8.10 & 7.93 & 161 & 16.4 & 0.28 \\
1q  & IC 5267  & no       & \nodata       & 7.85 & 7.68 & 157 & 23.1 & 0.32 \\
2a  & NGC 5643 & Sey 2    & \phn116       & 7.59 & 7.46 & 137 & 11.2 & 0.08 \\
2q  & NGC 4030 & no       & \nodata       & 7.68 & 7.60 & 125 & 16.5 & 0.10 \\
3a  & NGC 6300 & Sey 2    & 4272          & 7.29 & 7.20 & 134 & 11.1 & 0.14 \\
3q  & NGC 3368 & no       & \phn\phn\phn5 & 6.65 & 6.57 & 228 & 11.6 & 0.25 \\
4a  & NGC 6814 & Sey 1.5  & 4889          & 8.10 & 7.95 & \phn91 & 10.1 & 0.10 \\
4q  & NGC  628 & no       & \phn\phn\phn3 & 7.42 & 7.04 & 314 & 15.1 & 0.08 \\
5a  & NGC 7743 & Sey 2    & 1310 		  & 8.84 & 8.66 & \phn91 & \phn8.4 & 0.26 \\
5q  & NGC  357 & no       & \nodata       & 8.91 & 8.82 & \phn72 & 11.2 & 0.32 \\
\hline
\multicolumn{3}{l}{Seyfert Subsample}   & 2946      & 7.98/7.96      & 7.84/7.81      & 123/131 & 11.5/11.8 & 0.17/0.19 \\
  &    &                                & $\pm$2100 & $\pm$0.59/0.68 & $\pm$0.56/0.64 & $\pm$31/29 & $\pm$3.0/3.3 & $\pm$0.09/0.10 \\
\multicolumn{3}{l}{Quiescent Subsample} & \nodata   & 7.70/7.77      & 7.54/7.67      & 179/146 & 15.5/15.6 & 0.21/0.25 \\
  &    &                                & \nodata   & $\pm$0.82/0.93 & $\pm$0.84/0.92 & $\pm$94/65 & $\pm$4.8/5.6 & $\pm$0.12/0.10 \\
\hline
\multicolumn{3}{l}{Mean pair-wise difference} & \nodata & 0.28/0.18 & 0.30/0.15 & -57/-15 & -4.0/-3.8 & -0.04/-0.06 \\
  &    &                                      & \nodata & $\pm$0.37/0.34 & $\pm$0.47/0.37 & $\pm$104/53 & $\pm$2.4/2.7 & $\pm$0.05/0.04 \\

\enddata
\tablecomments{{\em Columns 1-3} give galaxy I.D.s, names, and classification.  The {\em X}-ray 2-10 keV luminosities given in {\em column 4} are from the literature: (1a) \citet{vasudevan10}, (2a) \citet{Lutz04}, (3a) \citet{Beckmann06}, (3q) and (4q) \citet{Ueda01}, (4a) \citet{Turner89}, (5a) \citet{panessa06}.  The luminosities given for 1a and 5a are corrected for these sources being Compton thick as discussed in \citet{vasudevan10} and \citet{panessa06}, respectively.  {\em Columns 5} and {\em 6} are {\em H}-band magnitudes as reported in the 2MASS Extended Source Catalog (\citealt{jarrett00}), where they are the magnitude within an elliptical aperture defined by the {\em Ks}-band 20 mag arcsec$^{-2}$ isophote and the total magnitude as derived from the isophotal aperture plus the integration of the surface brightness profile that extends from the isophotal aperture out to about 4 disk scale lengths, respectively.  {\em Columns 7} and {\em 8} are the galaxy radius, which is half the diameter $D_{25}$ of the B=25 mag arcsec$^{-2}$ isophote as reported in \citet{Martini03a}, and the distances used to translate to units of kpc are those given in Table \ref{tab:obs}.  The bulge-to-total flux ratios (B/T) estimated based on Hubble type (\citealt{laurikainen07}), which are given in Table \ref{tab:galprops}, are listed in {\em column 9}.  Average values of each these quantities for the two subsample are given in the bottom rows with the mean, e.g. $<0.5D_{25,active}>$, in the top row and the standard deviation in the row immediately below.  The means and standard deviations of the pair-wise differences, e.g. $<0.5D_{25,active}$-$0.5D_{25,quiescent}>$, for each quantity are given in the last two rows.  In both cases the first value is based on all 5 galaxy pairs and the second is excluding the fourth pair with NGC 628, which was excluded from the sample when considering stellar properties (see $\S$\ref{sec:628}).}

\end{deluxetable*} 


For the specific 5 Seyfert-quiescent galaxy pairs observed there are no statistically significant systematic differences in any of the five parameters on which the pairs were matched (e.g. there is no systematic difference in inclination; see Table \ref{tab:galprops}).  {\em H}-band luminosities and disk sizes were also compared for the pairs, and the subsamples as a whole, and some marginal differences are found.  The {\em H}-band magnitudes, which can be taken to be representative of galaxy mass, are found to be about 24\% (0.3 magnitudes; Table \ref{tab:galprops2}) fainter on average in the active galaxies.  If the pair containing NGC 628 is excluded, as is justified in $\S$\ref{sec:628}, then this difference drops to Seyferts being only 13\% fainter.  In addition, the galaxy sizes are about 23\% smaller in the active galaxy sample compared to the quiescent galaxies (Table \ref{tab:galprops2}).  In both cases the difference is only 1$\sigma$ or less, and it could explain as much as, but not more than, an approximately 25\% systematic difference in the subsamples due to the size and/or luminosity based on the sample selection.  Any difference in the individual bulge and disk components of the galaxy pairs is also likely to be limited to a similarly small difference since the bulge-to-total flux ratio is closely related to Hubble type (\citealt{laurikainen07}), which was used as a criterion in matching the galaxy pairs (see Table \ref{tab:galprops2}).  A comparison of the large-scale bar classification, which was not considered in the pair matching, shows that bars are found in all five Seyferts and two of the quiescent galaxies (NGC 3368 and NGC 357; \citealt{Martini03a}).  There is therefore also no significant subsample bias due to the presence of a large-scale bar, which is consistent with the lack of a correlation between such bars and Seyfert activity found in larger samples (e.g. \citealt{comeron08}).

\subsection{Observations and Reduction of Integral Field Data}
\label{sec:obs}
Near infrared integral field data was obtained with SINFONI (\citealt{eisenhauer03}, \citealt{bonnet04}) on the ESO Very Large Telescope (VLT).  The pixel scale of the R\sm4300 \kb\ spectra (approximately 1.95-2.45\mic) is 0\as.125$\times$0\as.25 and typically the field of view (FOV) measured is 8\as$\times$8\as. All observations were obtained in seeing-limited mode with mean point spread function (PSF) full-width-half-maximum (FWHM) of 0.55$\pm$0.05\as~or 54$\pm$24 pc.  This excludes NGC 4030, for which all data is of worse quality due to rapidly degrading seeing conditions and is estimated to be around 1\as (or 132 pc) based on the ESO seeing monitor, and may partially explain the lack of \htwo~detection.  Integration times range from 50-140 minutes per galaxy.  See Table \ref{tab:obs} for the details of each dataset.

\begin{deluxetable}{clcccccc}
\tabletypesize{\small}
\tablecaption{Summary of Observations \label{tab:obs}} 
\tablewidth{0pt}
\tablehead{
\colhead{I.D.} &
\colhead{Galaxy} &
\colhead{D} &
\colhead{Ref.\tablenotemark{b}} &
\colhead{pc/\as} &
\colhead{T$_{int}$} &
\multicolumn{2}{c}{PSF FWHM} \\
\colhead{} &
\colhead{} &
\colhead{(Mpc)} &
\colhead{} &
\colhead{} &
\colhead{(min)} &
\colhead{(\as)} &
\colhead{(pc)}  \\
}
\startdata

1a & NGC 3227 & 21.1 & 1 & 102    & \phn50  & 0.55 & 56 \\
1q & IC 5267  & 30.3 & 2 & 147    & 140     & 0.61 & 90 \\
2a & NGC 5643 & 16.9 & 3 & \phn82 & 140     & 0.49 & 40 \\
2q & NGC 4030\tablenotemark{a} & 27.2  & 4 & 132 & \phn50\tablenotemark{a}  & 0.66\tablenotemark{a} & 87\tablenotemark{a} \\
3a & NGC 6300 & 17.1 & 4 & \phn83 & 140     & 0.48 & 40 \\
3q & NGC 3368 & 10.5 & 5 & \phn51 & \phn40  & 0.58 & 30 \\
4a & NGC 6814 & 22.8 & 3 & 111    & 140     & 0.51 & 57 \\
4q & NGC  628 & \phn9.9 & 6 & \phn48 & 100     & 0.59 & 28 \\
5a & NGC 7743 & 19.2 & 7 & \phn93     & 140     & 0.54 & 50 \\
5q & NGC  357 & 32.1 & 3 & 156    & \phn90  & 0.62 & 97 \\

\enddata
\tablenotetext{a}{Observing conditions degraded rapidly during the acquisition of the data for NGC 4030 and this PSF FWHM measurement reflects a lower limit.  The actual PSF FWHM during the majority of the NGC 4030 data is estimated to be closer to 1\as~based on ESO seeing monitor at Paranal.}
\tablenotetext{b}{References for distances: (1) \citealt{terry02}, (2) \citealt{willick97}, (3) \citealt{tully88}, (4) \citealt{tully09}, (5) \citealt{freedman01}, (6) \citealt{olivares10}, (7) \citealt{jensen03}.}

\end{deluxetable} 


The spatial resolution is determined from PSF calibration frames taken immediately before or after the science frames.  For a few of the Seyfert galaxies the PSF can also be estimated from the data themselves.  This is achieved by applying the fact that the intrinsic equivalent width of the CO 2-0 bandhead is expected to fall within a limited range (\citealt{oliva95}, \citealt{forster00}, \citealt{Davies07}).  The observed equivalent width can therefore be used to determine the dilution of the stellar feature due to the AGN continuum emission (see $\S$\ref{sec:agn} and $\S$\ref{sec:cubefits} for details), which, at this resolution, is a point source (predicted sizes are $<$ 2 pc; \citealt{jaffe04}, \citealt{tristram07}, \citealt{raban09}).  PSF FWHM values determined directly from the data are consistent with those determined from the calibration frames to within 0.02\as, which is consistent with expected variation in the ambient seeing with time between the acquisition of the data and calibration frames.

The data were reduced using the SINFONI custom reduction package SPRED (\citealt{abuter06}), which includes all of the typical reduction steps applied to near infrared spectra with the additional routines necessary to reconstruct the data cube.  After background subtraction, which is performed with sky frames obtained interspersed with the on-source exposures, the data are flat-fielded and corrected for dead/hot pixels.  Telluric correction and flux calibration, which is accurate to 10\%, are carried out using A- and B-type stars.  Residuals from the OH line emission are minimized using the methods outlined in \citet{davies07b}.

\subsection{Confirmation of AGN classification}
\label{sec:agn}

\begin{figure}[t]	
\epsscale{1.0}
\plotone{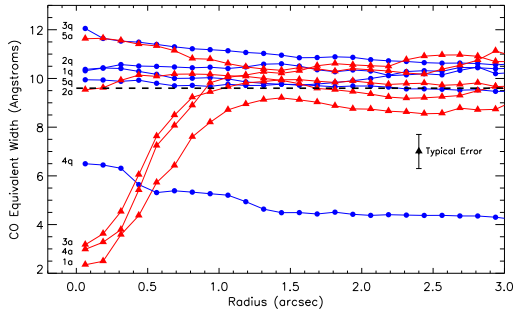}
\caption[]{Azimuthal average of the CO equivalent width.  Significant dilution is seen in only three galaxies, all Seyferts.  Triangles are Seyfert galaxies while circles are quiescent galaxies, and all curves are labeled according to those defined in Table \ref{tab:galprops}.  The horizontal dashed line represents the lower bound on an undiluted CO bandhead equivalent width (see $\S$ \ref{sec:agn} for details).  A typical error bar, representative of the standard deviation of the equivalent width of all pixels at each annulus, is shown on the right. \label{fig:co_eqw}}
\end{figure}

The luminosity of the non-stellar AGN continuum is estimated in each galaxy by measuring the equivalent width (EQW) of the $^{12}$CO(2,0) \lam2.2935 bandhead.  The intrinsic equivalent width of this absorption is constant for a broad range of star formation histories and ages and a deviation by more than 20\% of the contant value of 12 \AA, i.e. lower than 9.6 \AA, implies dilution by the AGN continuum (\citealt{oliva95}, \citealt{forster00}, \citealt{Davies07}).  The absence of such dilution does not rule out the possibility of an AGN buried in a very optically thick material (i.e. a Compton thick AGN), but the presence of dilution can be taken as verification of an AGN continuum.  Alternatively a very young ($<$10 Myr) stellar population can cause similar dilution, but there is no indication of such a population in our sample galaxies.  

\begin{figure}[t]	
\epsscale{1.2}
\plotone{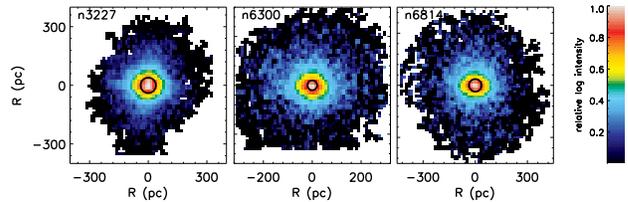}
\caption[]{Maps of non-stellar continuum in three galaxies (NGC 3227, NGC 6300, and NGC 6814) where significant dilution of the CO bandheads is seen due to an AGN continuum.  The central circle represents the radius at which the dilution factor is 2.  In all maps North is up and East is to the left.  \label{fig:NS_maps}}
\end{figure}

As shown in Fig. \ref{fig:co_eqw}, there is significant dilution of the CO equivalent width in three of the Seyfert galaxies: NGC 3227 (I.D. 1a as defined in Table \ref{tab:galprops}), NGC 6300 (3a), and NGC 6814 (4a).  A 2D map of the AGN continuum, calculated assuming the non-stellar continuum is equal to the total continuum times (1 - EQW$_{observed}$/EQW$_{intrinsic}$), is shown in Fig. \ref{fig:NS_maps} for each of these galaxies.  There is no indication of significant dilution from an AGN continuum in the remaining two active galaxies, NGC 5643 (2a) and NGC 7743 (5a), despite their classification based on optical line ratios as Seyfert 2 galaxies (\citealt{phillips83} and \citealt{ho97a}, respectively).  However, there is a weak downturn in the CO equivalent width of NGC 5643 (2a), which reaches the 9.6\AA~cut off for AGN dilution at the very center.  The lack of significant dilution of the CO bandheads in these two galaxies suggests that they have lower levels of AGN activity in comparison to the others and/or they are Compton thick.  In both cases there is evidence of their being Compton thick (as discussed below), and for NGC 7743 some studies find that it is consistent with a low luminosity AGN (e.g. \citealt{Gonzalez09}).  As will be discussed in $\S$\ref{sec:global}, these two galaxies still exhibit nuclear molecular gas and stellar properties that set them apart from their paired quiescent galaxies following the same trends seen in the other Seyfert galaxies.  In all quiescent galaxies there is no indication of an AGN continuum, confirming their non-AGN classification with a relatively constant CO equivalent width across the field of view, including at the center as defined by the peak of the {\em K}-band luminosity.  As will be discussed in $\S$\ref{sec:628}, NGC 628 has a CO equivalent width significantly lower than all other galaxies in the sample, which can be explained by a relatively small contribution of older stars.

The AGN contribution in each galaxy is also assessed by gathering available 2-10 keV X-ray luminosities from the literature (Table \ref{tab:galprops2}).  With the exception of Compton thick sources, AGN are expected to produce a significant level of 2-10 keV emission and it can thus be used as an indicator of such activity.  Based on available measurements, it is found that the selected active galaxies have 2-3 orders of magnitude higher X-ray luminosities compared to the quiescent galaxies in the sample (three of the quiescent galaxies were not detected in surveys in which they were included).  This difference in X-ray luminosities suggests that our selected AGN are indeed significantly more active than the selected quiescent galaxies, and are consistent with a Seyfert classification.  There is evidence of significant obscuration in three of the Seyfert galaxies: NGC 3227 (\citealt{vasudevan10}), NGC 5643 (\citealt{Lutz04}, \citealt{guainazzi04}, \citealt{shu07}), and NGC 7743 (\citealt{panessa06}).  For two of these Compton thick sources, NGC 3227 and NGC 7743, an estimate of the intrinsic 2-10 keV emission (corrected for obscuration) is available and listed in Table \ref{tab:galprops2}.  Also of note is that the three galaxies with the strongest non-stellar AGN continua are also those with the highest X-ray luminosities, that NGC 5643 (2a), which has only a hint of AGN continuum based on dilution of the CO bandheads, has moderate observed X-ray luminosities and is known to be highly obscured, and that NGC 7743 (5a), which shows no indication of an AGN continuum, has an observed X-ray luminosity comparable to the brightest quiescent galaxy (\citealt{Gonzalez09}; i.e. before correction for obscuration).

\section{Characterization of Nuclear Molecular Gas and Stars}
\label{sec:maps}

\subsection{2D Distribution and Kinematics}
\label{sec:cubefits}

\begin{figure}[b]	
\epsscale{1.2}
\plotone{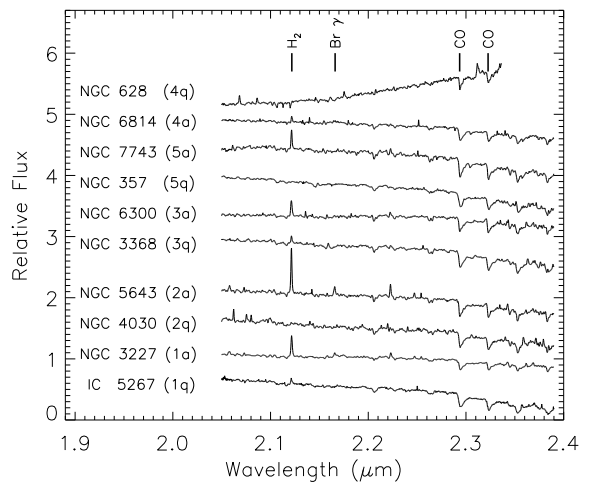}
\caption[]{Spectra integrated over a 8"x8" field of view arbitrarily shifted for clarity. The summed spaxels for each galaxy have been corrected to remove the rotational velocity based on a smoothed mapped of the stellar velocity field and each spectrum has been deredshifted. \label{fig:intspec}}
\end{figure}

\begin{figure*}[t]	
\epsscale{1.0}
\plotone{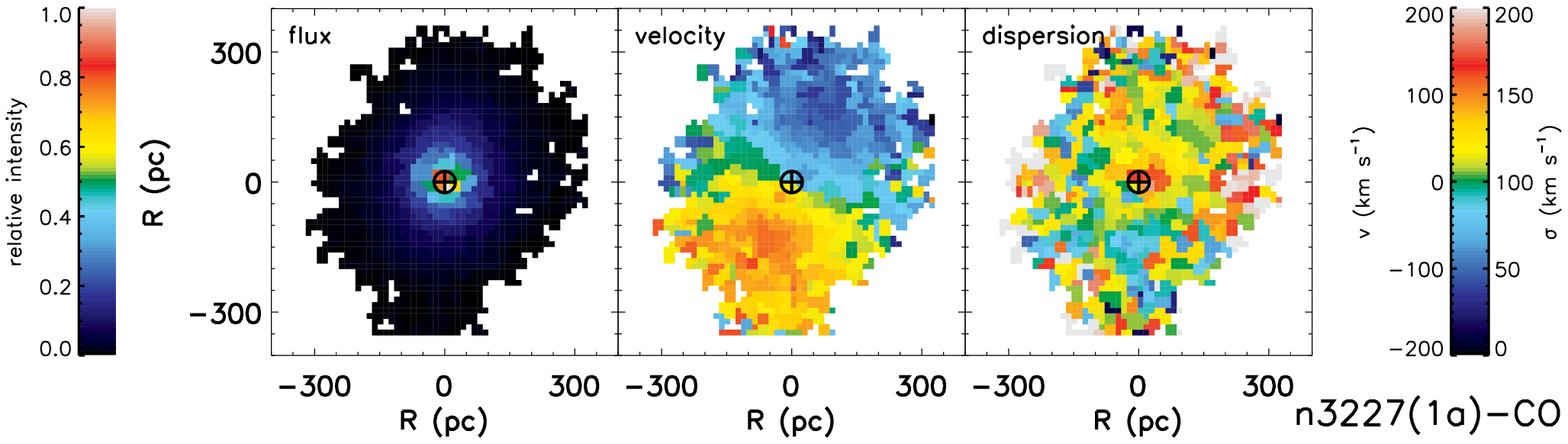}
\plotone{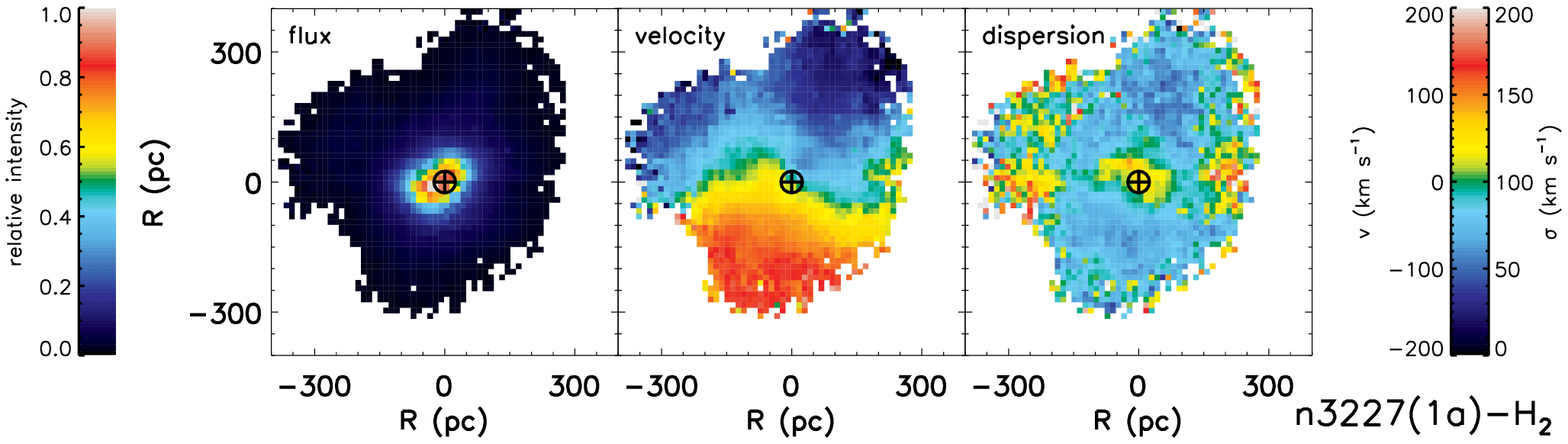}
\caption[]{Pair 1: NGC 3227 (1a) 2D maps of CO absorption (stars) and \htwo~emission (molecular gas).  Maps are as labeled from left to right: observed flux, velocity, and velocity dispersion.  2D maps of CO are in the top row and \htwo~in the bottom row, as indicated in the lower right of each set along with the galaxy name and pair identification.  In all maps north is up and east is to the left.  \label{fig:3227_maps}}
\end{figure*}

\begin{figure*}[t]	
\epsscale{1.0}
\plotone{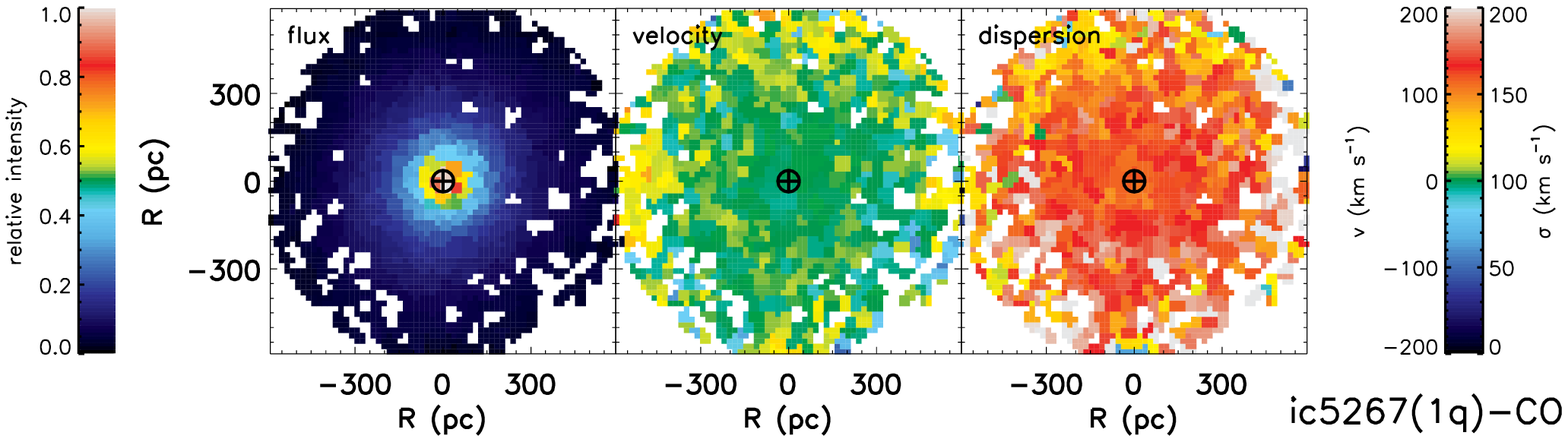}
\plotone{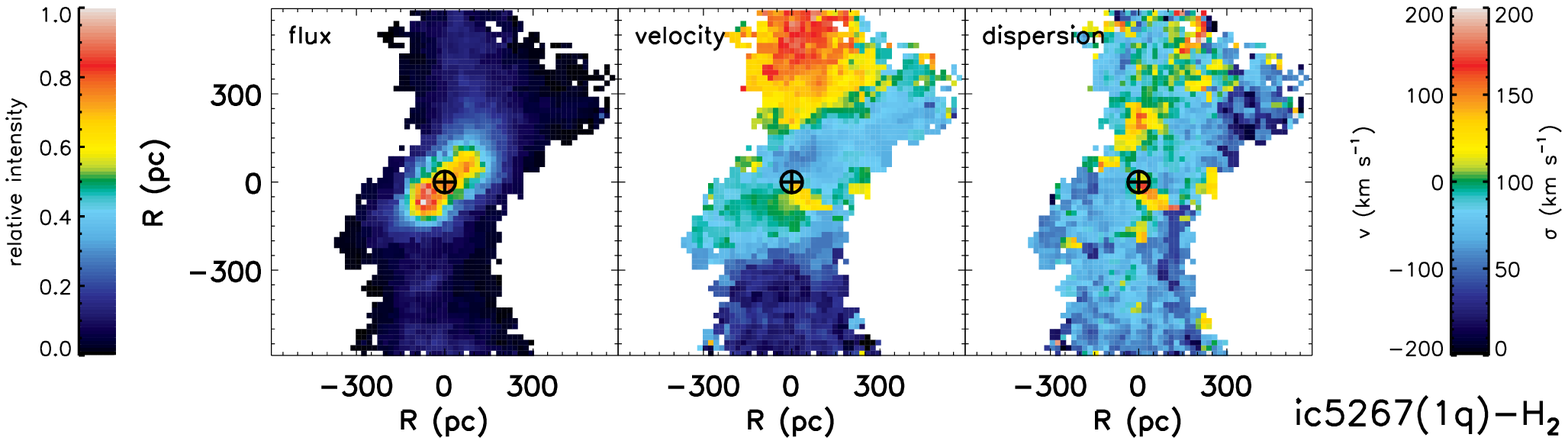}
\caption[]{Pair 1: IC 5267 (1q) 2D maps.  See Fig. \ref{fig:3227_maps} for details.  \label{fig:5267_maps}}
\end{figure*}
In all 10 galaxies the first two CO bandheads at $\lambda$2.2935, 2.3226 \mic~are detected well enough to reliably measure the velocity and velocity dispersion across a large fraction of the 8\as$\times$8\as~field of view.  \htwos~$\lambda$2.1218 \mic~emission is detected in 7 of the galaxies, with all 3 of those with non-detections being quiescent galaxies.  Two of these non-detections are galaxies for which Martini et al. (2003a) find no detectable nuclear dust structure, which is known to correlate with low molecular gas content (e.g. \citealt{young91}, \citealt{dumas10}).   Spectra of each galaxy integrated over the measured field-of-view are shown in Fig. \ref{fig:intspec}.

\begin{figure*}[!ht]	
\epsscale{1.0}
\plotone{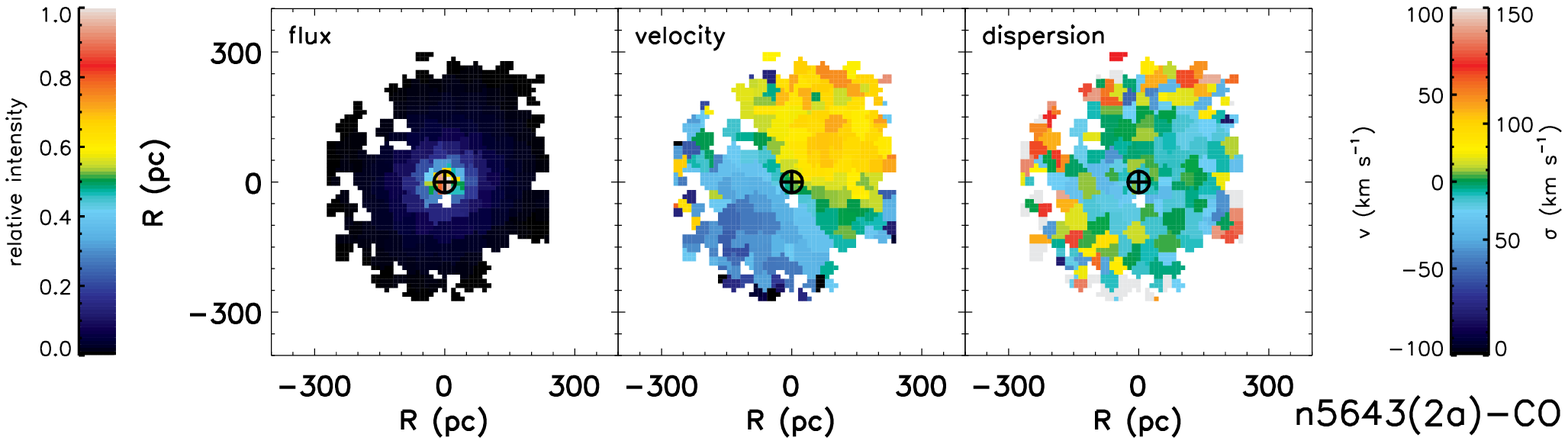}
\plotone{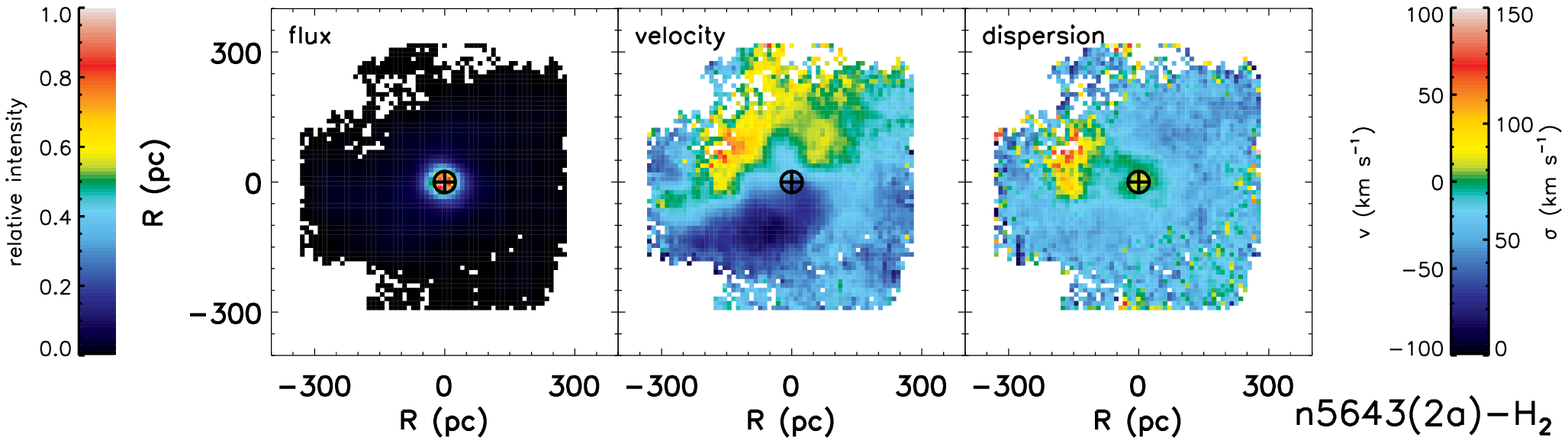}
\caption[]{Pair 2: NGC 5643 (2a) 2D maps.  See Fig. \ref{fig:3227_maps} for details. \label{fig:5643_maps}}
\end{figure*}

\begin{figure*}[!ht]	
\epsscale{1.0}
\plotone{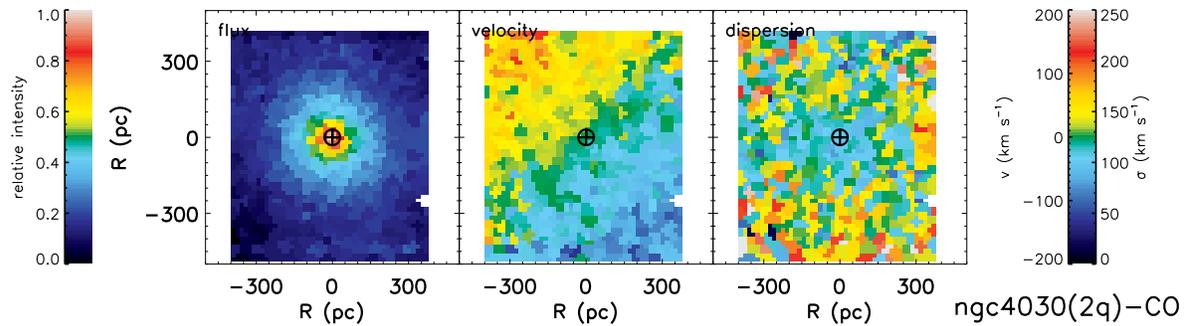}
\caption[]{Pair 2: NGC 4030 (2q) 2D maps.  See Fig. \ref{fig:3227_maps} for details. \label{fig:4030_maps}}
\end{figure*}

The 2D flux distribution, velocity, and velocity dispersion (\sig) of the stellar and gas kinematics have been extracted by fitting the CO bandheads and \htwo~emission, respectively.  Before extraction of the kinematics the data cubes were spatially binned to a minimum SNR with the Voronoi binning technique of \citet{Cappellari03}.  This was done to increase the SNR in outer radii where otherwise the signal would be too low for reliable line profile characterization.  A target SNR of 10 and 5 per bin was used for all stellar and molecular gas maps, respectively.  This level of binning maintained as high a spatial resolution as possible within the central 200 pc while still improving the SNR in outer radii.  
\begin{figure*}[!ht]	
\epsscale{1.0}
\plotone{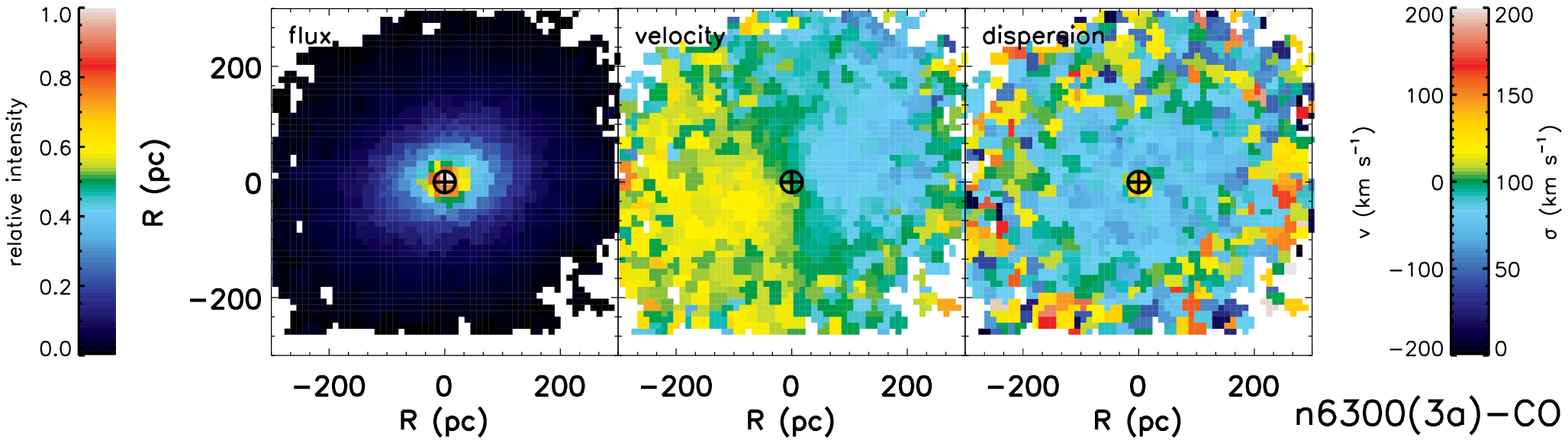}
\plotone{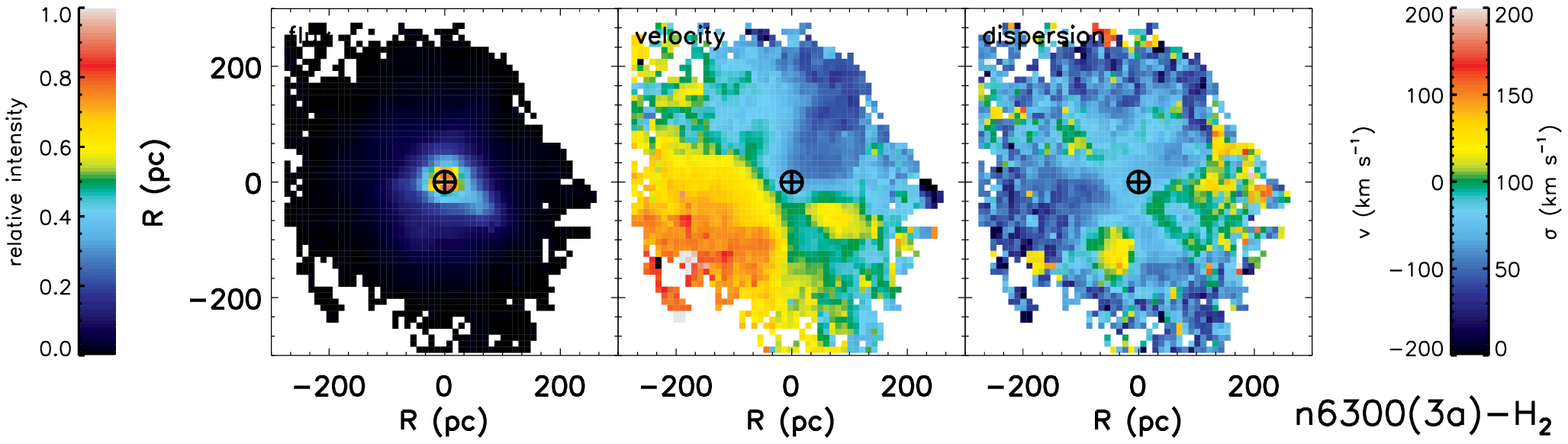}
\caption[]{Pair 3: NGC 6300 (3a) 2D maps.  See Fig. \ref{fig:3227_maps} for details. \label{fig:6300_maps}}
\end{figure*}

\begin{figure*}[!ht]	
\epsscale{1.0}
\plotone{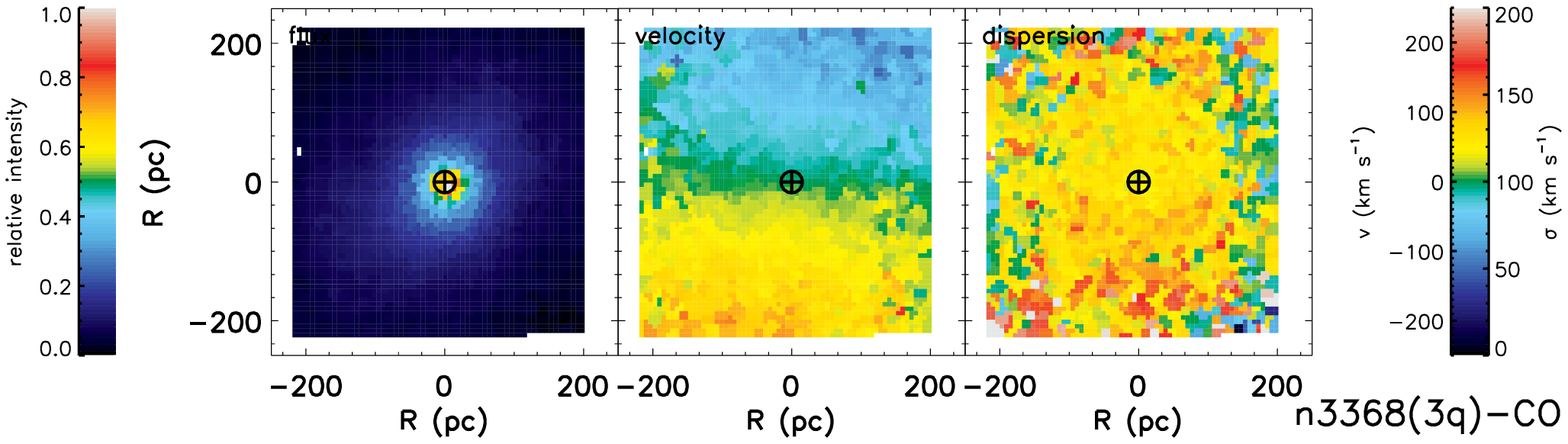}
\plotone{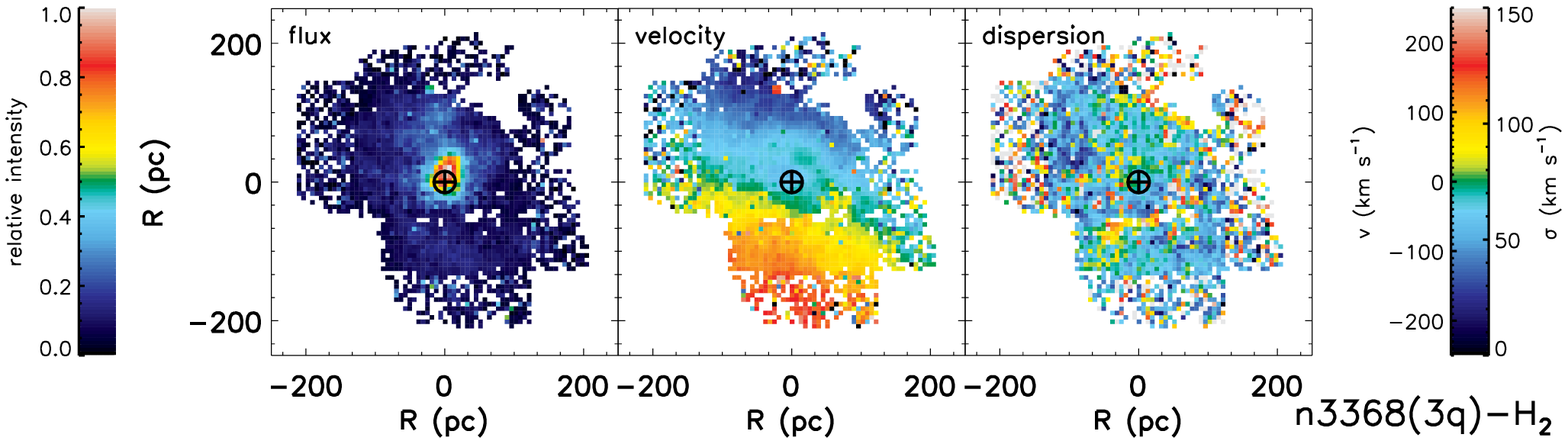}
\caption[]{Pair 3: NGC 3368 (3q) 2D maps.  See Fig. \ref{fig:3227_maps} for details.  Note that for this particular galaxy the spatial coverage for \htwo~ is dictated by the SNR which was only high enough for reliable emission line characterization in the central overlapping region of the four science frames obtained.  The SNR of the CO bandheads was high enough that line characterization out to the edge of each science frame was possible and therefore wider spatial coverage was achieved.
\label{fig:3368_maps}}
\end{figure*}

\begin{figure*}[!ht]	
\epsscale{1.0}
\plotone{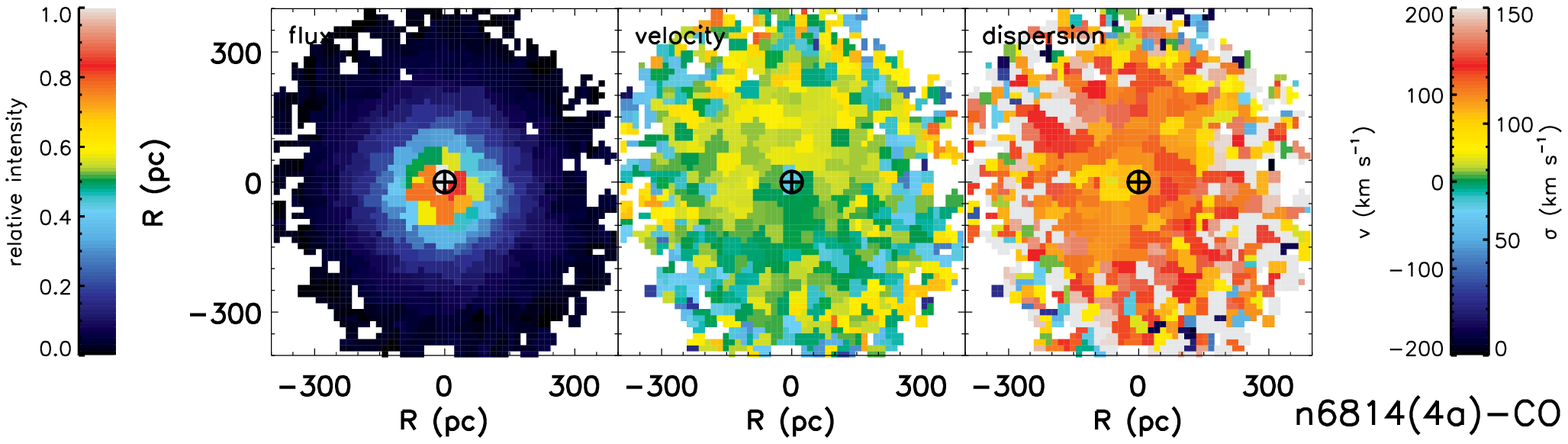}
\plotone{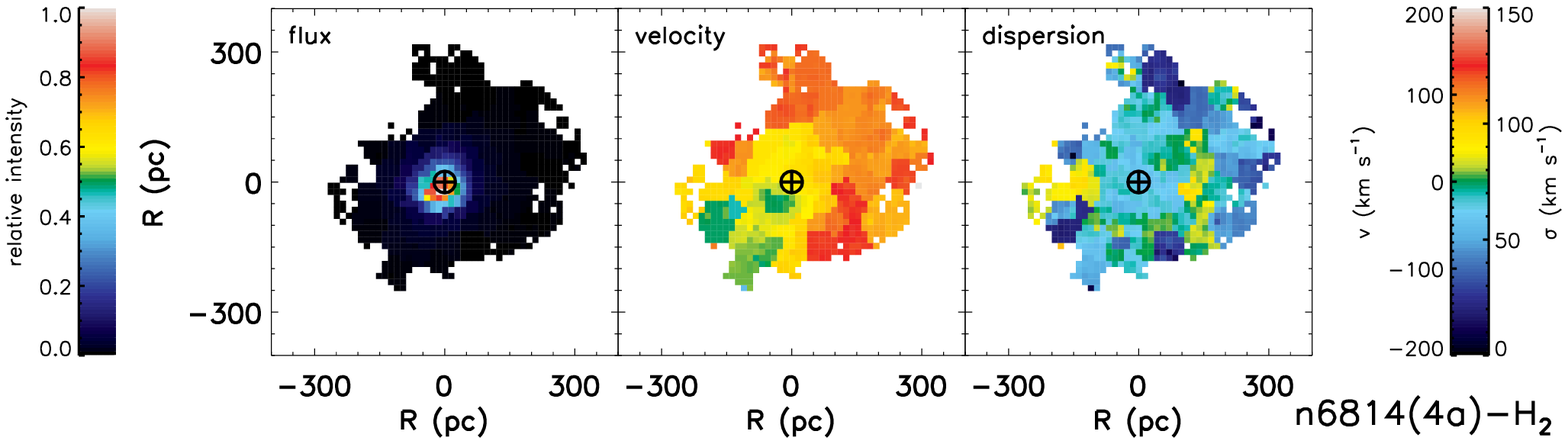}
\caption[]{Pair 4: NGC 6814 (4a) 2D maps.  See Fig. \ref{fig:3227_maps} for details. \label{fig:6814_maps}}
\end{figure*}

\begin{figure*}[!ht]	
\epsscale{1.0}
\plotone{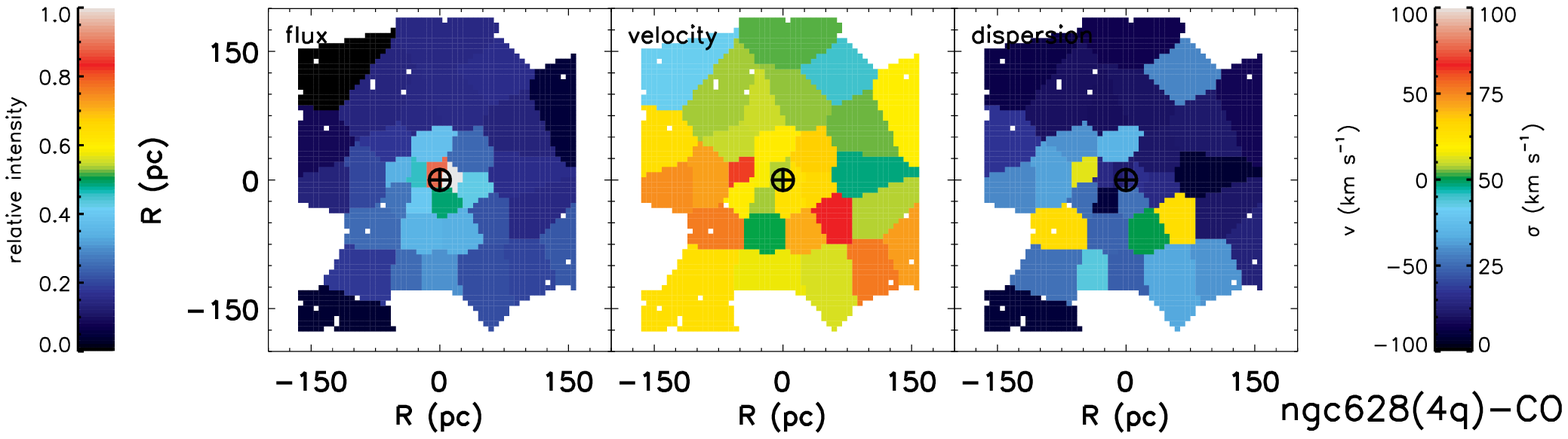}
\caption[]{Pair 4: NGC 628 (4q) 2D maps.  See Fig. \ref{fig:3227_maps} for details. \label{fig:628_maps}}
\end{figure*}

\begin{figure*}[!ht]	
\epsscale{1.0}
\plotone{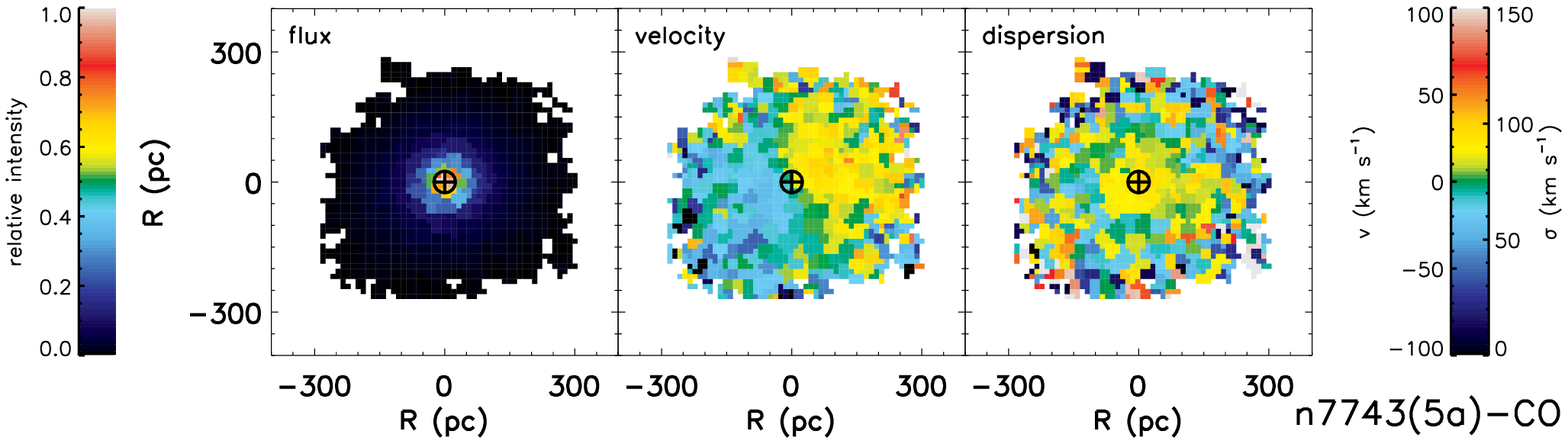}
\plotone{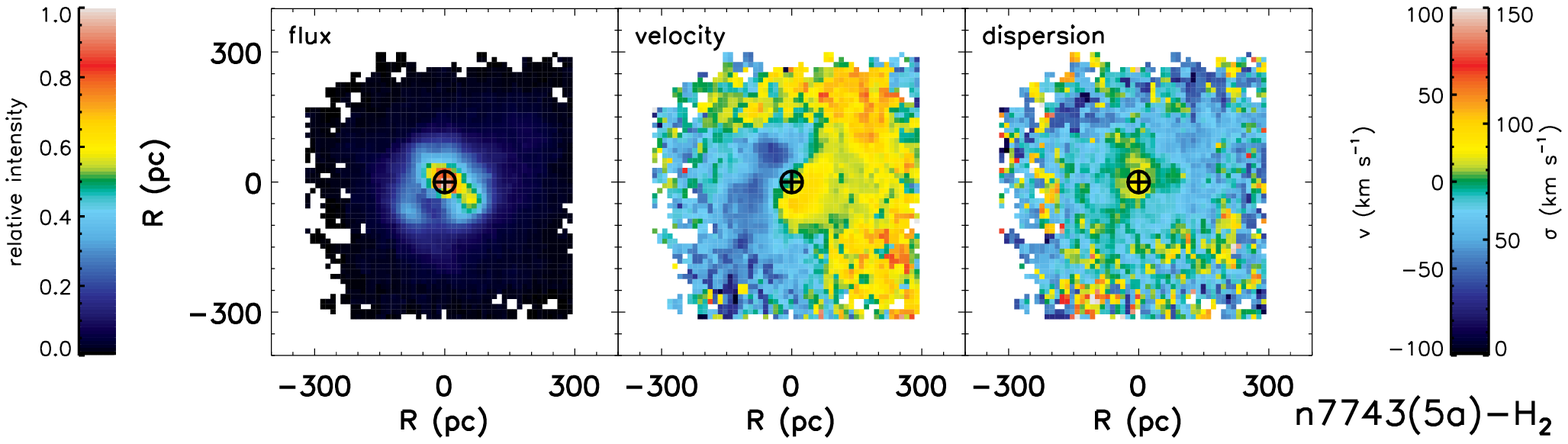}
\caption[]{Pair 5: NGC 7743 (5a) 2D maps.  See Fig. \ref{fig:3227_maps} for details. \label{fig:7743_maps}}
\end{figure*}

\begin{figure*}[!ht]	
\epsscale{1.0}
\plotone{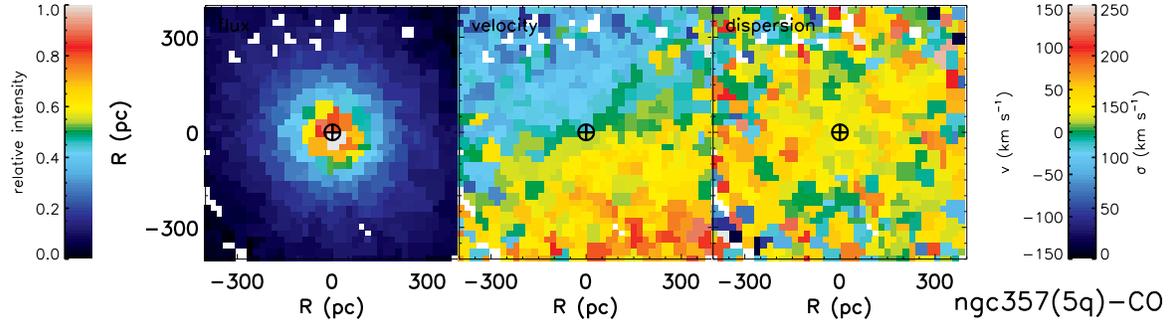}
\caption[]{Pair 5: NGC 357 (5q) 2D maps.  See Fig. \ref{fig:3227_maps} for details. \label{fig:357_maps}}
\end{figure*}

The line profile of the \htwo~emission is extracted using our code LINEFIT described in \citet{davies11} (see also \citealt{forster09}).  Briefly, this code fits the emission line in the spectrum of each spatial element, or spaxel, with an unresolved line profile (e.g. a sky line) convolved with a Gaussian, as well as a linear function to the continuum.  Similarly, the $\lambda$2.2935, 2.3226 \mic~CO bandheads were simultaneously fit with a broadened stellar template at each spaxel (except in NGC 628 where only the first CO bandhead was fit due to the limited spatial region over which the signal-to-noise ratio (SNR) of the second bandhead is adequate for a robust measurement).  Two late-type stellar standards were obtained for this purpose as part of our observing program and both are found to be well matched with our galaxy spectra.  For the Seyfert galaxies a region just blueward of the second CO bandhead (from 2.318-2.323 \mic) was excluded from the fit to avoid any contamination from the [Ca VIII] $\lambda$2.3213 \mic~coronal line which is present in at least 4 of the 5 active galaxies (see $\S$ \ref{sec:ionized}).  All reported velocity dispersion values are intrinsic values, i.e. an instrumental broadening of 35 \kms~has been accounted for.  The uncertainty of fits to both the CO bandheads and the \htwo~emission are estimated using Monte Carlo techniques by refitting with added noise of the same statistics as the data.  This is done 100 times, and the standard deviations of the best-fit parameters are used as the uncertainties.  Typically the uncertainties in velocity and velocity dispersion for both the stars and gas are on the order of 5-15 \kms.  Bins with poor quality fits (generally SNR below 2 or kinematic errors greater than 30 km/s) were also masked.  It is on these masked binned maps that all subsequent results are based for both the CO bandheads and \htwo.  The resulting 2D maps of the stellar and molecular gas distribution and kinematics are shown in Figures \ref{fig:3227_maps} through \ref{fig:357_maps}.

The stellar luminosities are corrected for AGN contamination based on the dilution of the CO absorption equivalent width.  As discussed in $\S$ \ref{sec:agn}, to within 20\% the intrinsic {\it K}-band CO equivalent width is 12 \AA, and a deviation from this value represents dilution from the non-stellar (e.g. AGN) continuum (e.g. \citealt{Davies07}).  The stellar fraction of the continuum is therefore equivalent to the ratio of the observed CO equivalent width to the intrinsic value, and all stellar luminosities reported are corrected using this estimate.

The kinematic center of the galaxy is taken to be that determined from doing a 2D Gaussian fit to the non-stellar (i.e. AGN) continuum, when available, or the {\em K}-band continuum, which is dominated in all galaxies by the stellar continuum.  For the three galaxies with strong non-stellar continua, the centers based on the non-stellar continua and the total {\em K}-band continua are consistent within the measurement error of a tenth of a pixel.

\subsection{The Unique Case of NGC 628}
\label{sec:628}

Compared to the rest of the observed sample the stellar attributes of NGC 628 are unique in a number of critical ways that warrants excluding this galaxy from our comparative analysis of the stellar properties (but not the \htwo~properties) in our two subsamples.  This is justified by the fact that in this galaxy: (1) the spectral slope is red in contrast to the blue slope seen in all other galaxies (see Fig. \ref{fig:intspec}), (2) its stellar luminosity is more than 40 times lower than the other galaxies within the field of view measured (out to a radius of 150 pc), (3) the CO bandhead velocity dispersion is 30 \kms~or less throughout the field of view measured, which is 4.2 times lower than the rest of the sample (5.0 and 3.5 times lower than the quiescent and Seyferts galaxy subsamples, respectively), and (4) the equivalent width of the CO bandheads is around 5\AA~with a peak value at the center of 6.8\AA, which is well below that expected if the stellar light is dominated by the late-type stars typically present on the 0.1-1 kpc scales observed.  Each of these measurements is unchanged when considering a fit to the first two CO bandheads rather than just the first bandhead on which our presented results are based as discussed in $\S$ \ref{sec:cubefits}.  The low stellar velocity dispersion, equivalent width, and luminosity could suggest that the contribution from the bulge, at least on the 0.1-1kpc scales observed, is very weak or absent such that the near-IR continuum is not dominated by the typical late-type stars of a ``classical'' bulge.  In this scenario, the very low stellar luminosity can also explain the redness of the spectral slope since even a small amount of warm/hot dust emission, which could be (and likely is) present in any of the other sample galaxies, will have a much more significant impact on the total continuum of NGC 628.  Any dust emission would dilute the CO bandheads produced by the relatively few late-type stars present on the observed scales in this galaxy. In addition, to heat the dust on spatially extended ($\sim$100 pc) scales, one requires a population of young ($<$10 Myr) stars. The near-IR continuum of these young stars, while weak, would further reduce the depth of the CO bandheads

\begin{figure}[t]	
\epsscale{1.2}
\plotone{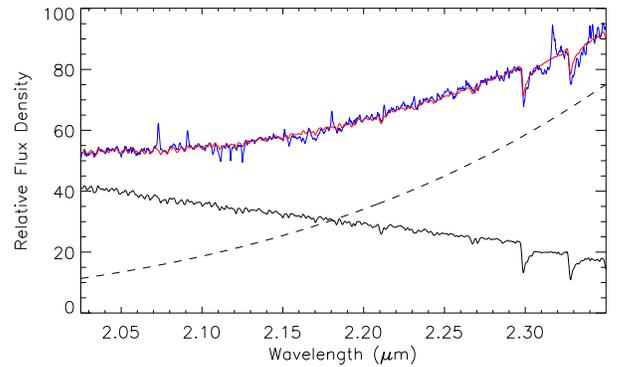}
\caption[]{Fit to the integrated spectrum of NGC 628.  The fit (red curve) overlaid on the integrated spectrum (blue curve) consists of a sum of two components: late-type stars (solid black curve) and 370 K dust emission (dashed black curve; assumes an emissivity $Q_{v}$=1).  \label{fig:628specfit}}
\end{figure}

To test the above hypothesis the integrated spectrum of NGC 628 is fit with a model composed of a composite template stellar spectrum and a dust component assuming a simple black body spectrum (i.e. emissivity $Q_v$=1).  As shown in Fig. \ref{fig:628specfit}, a fit to the spectrum of NGC 628 is consistent with very little stellar continuum generated by late-type stars and is instead dominated by a contribution from warm/hot dust emission.  The best-fit stellar template, which was generated from a stellar library (with no reddening applied), consisted of a combination of only late-type supergiants and no F- or G-type stars.  The best-fit dust temperature was 370 K, which results in a stellar fraction of 0.33 at 2.29 \mic.  The observed low CO bandhead equivalent width of 5\AA~can also be used to roughly estimate the stellar fraction since, as discussed, its value should intrinsically be about 12\AA.  This approach gives a stellar fraction of 0.42 (=5\AA/12\AA.), which is reasonably consistent with the fraction derived from the spectral fit.  The spectrum of NGC 628 is thus dominated by warm/hot dust emission (potentially heated by a young stellar population which does not contribute to the near-IR emission) because of the lack of a significant contribution from late-type stars.  Supporting this picture, \citet{zou11} find evidence for the previously presumed spheroid to actually be composed of a disk-like stellar component, a nucleus, and nuclear spiral plus ring structures with young stellar populations.  Interestingly, this conclusion suggests that the observed unique near-IR photometric and kinematic characteristics might be a useful way to identify galaxies for which the spheroid is a mis-classified disk-like structure.

\section{Differences in Global Properties Within a Radius of 250 pc}
\label{sec:global}

\begin{figure*}[t]	
\epsscale{1.0}
\plotone{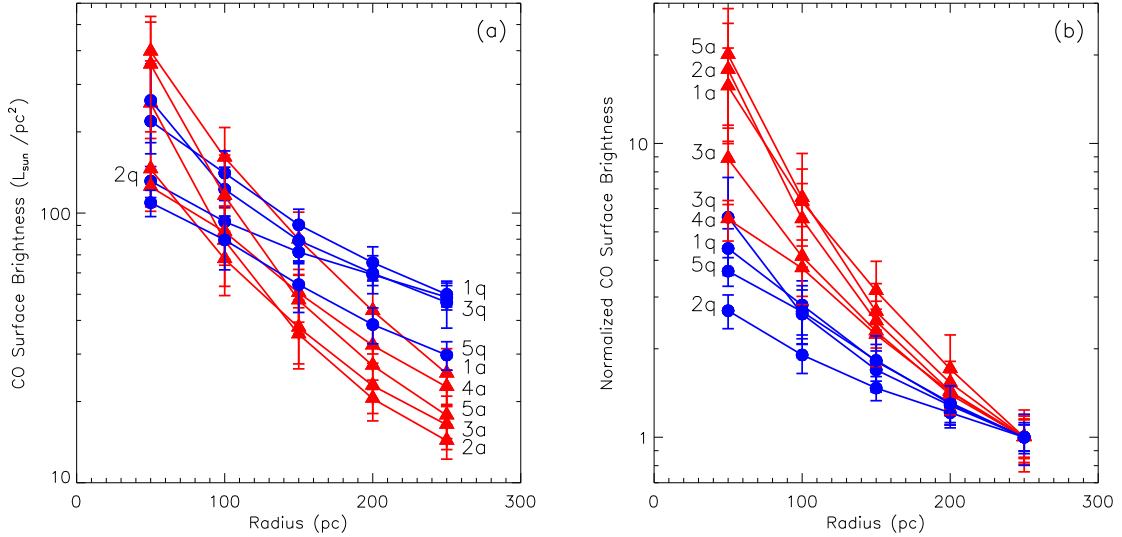}
\caption[]{Mean stellar luminosity as a function of radius (a) intrinsic values and (b) normalized to 250 pc.  Error bars are the standard deviations of measurements within each aperture.  Triangles represent Seyfert galaxies while circles are quiescent galaxies.  The label on the curve for each galaxy is as given in Table \ref{tab:galprops}. \label{fig:co_azlum}}
\end{figure*}

\begin{figure*}[t]	
\epsscale{1.2}
\plotone{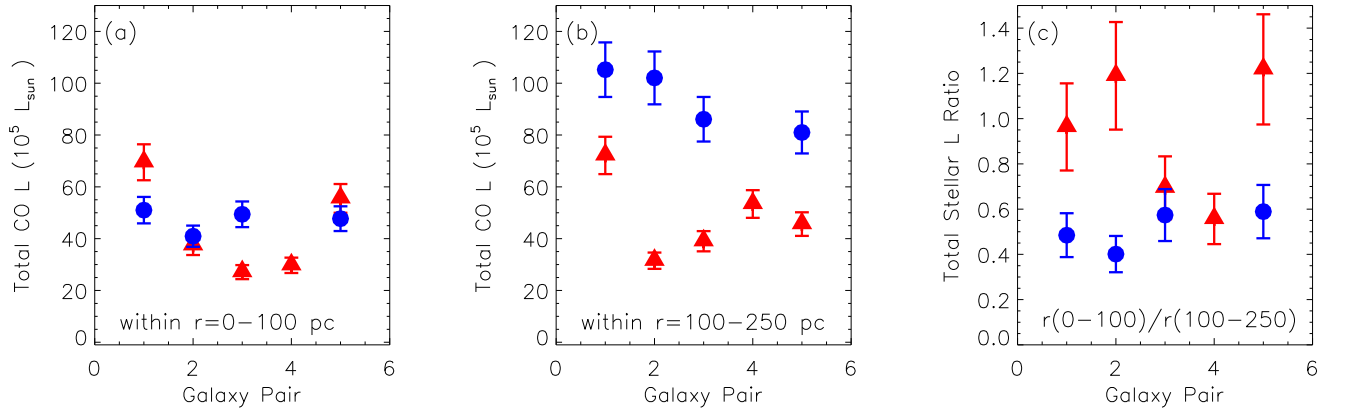}
\caption[]{Total CO luminosity in apertures of radius (a) 0 to 100 pc and (b) 100 to 250 pc and (c) the ratio of these two.  Triangles represent Seyfert galaxies while circles are quiescent galaxies.  Error bars represent the 10\% systematic uncertainty of the flux calibration.  The probability that the two subsamples are drawn from different parent populations based on the  r=100-250 pc aperture is 93.8\%. \label{fig:pair_colum}}
\end{figure*}

\begin{figure*}[t]	
\epsscale{0.55}
\plotone{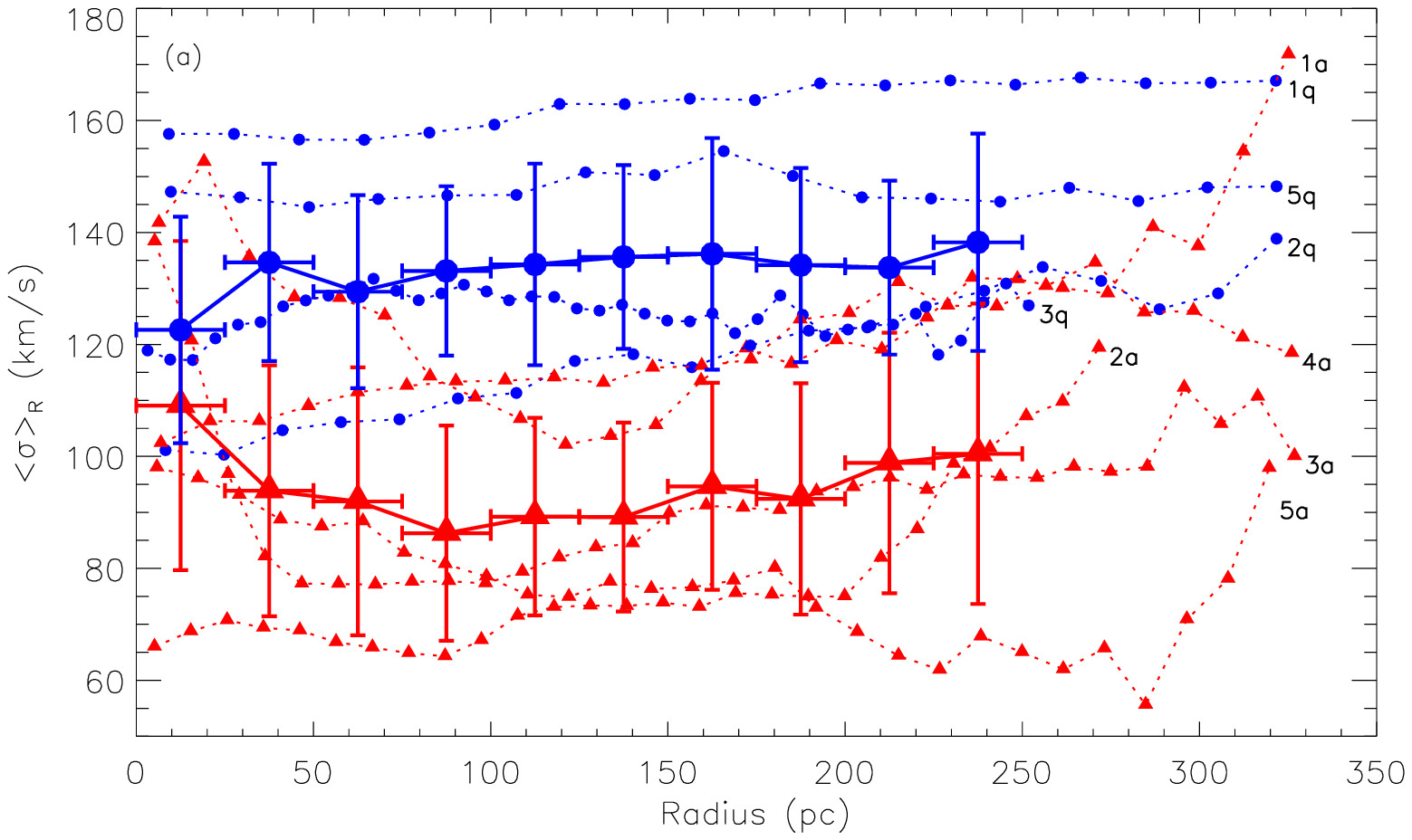}
\plotone{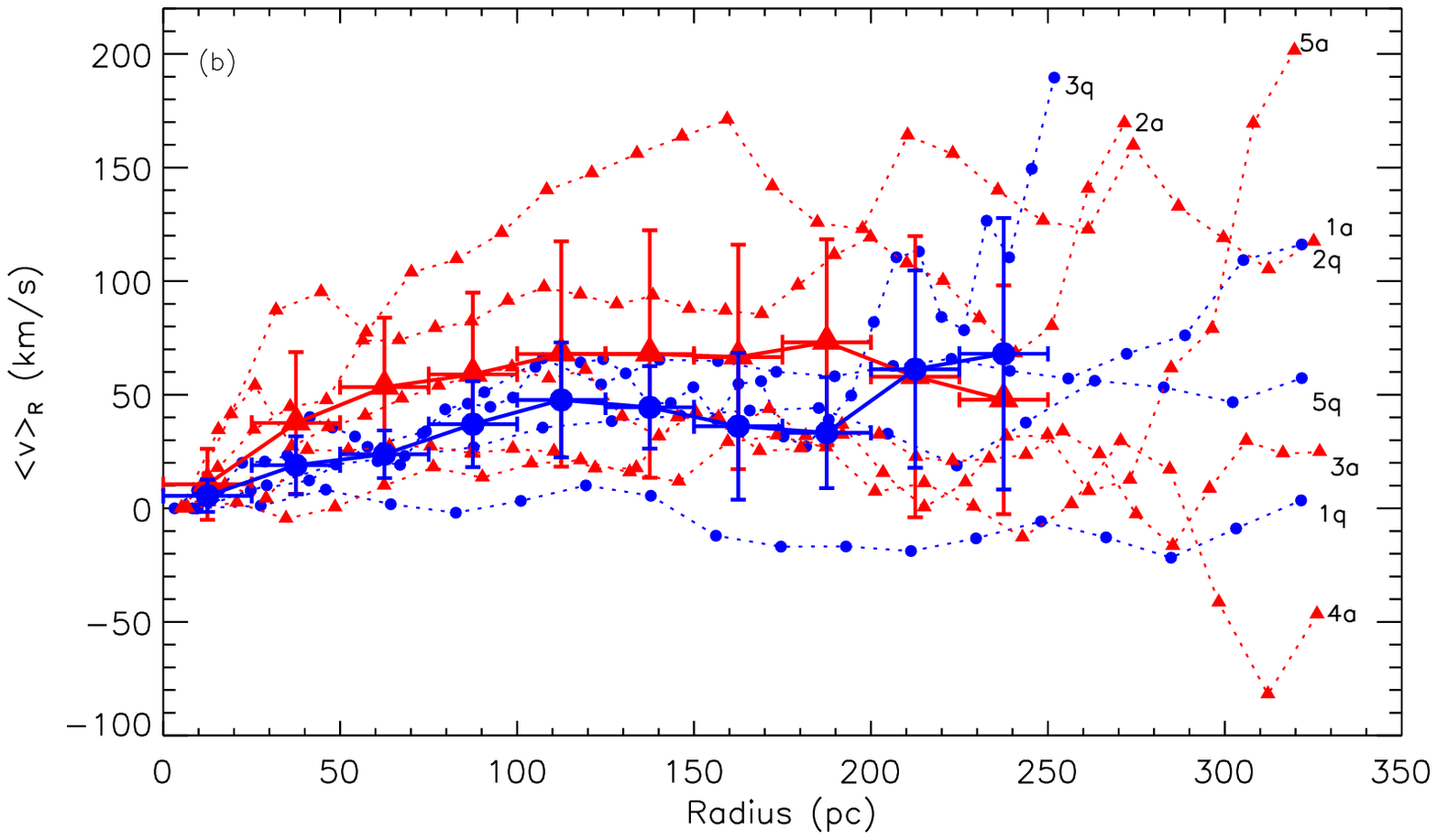}
\caption[]{Radially averaged kinematical properties of CO: (a) mean CO velocity dispersion as a function of radius and (b) inclination corrected rotational velocity.  The CO velocity dispersion is the mean of pixels at each radius bin uniformly weight (rather than luminosity weighted), and velocities have been corrected for inclination based on the large scale host galaxy inclination angles reported in \citet{Martini03a}.  The negative velocities beyond 150 pc for IC 5267 (1q), which are unphysical, are consistent within the error of the measurement with no detected rotation in this galaxy.  The larger symbols and solid curve represent the mean of each subsample and the errors bars are the standard deviation of the mean at that radius and the radial bin size.  Triangles represent Seyfert galaxies while circles are quiescent galaxies.  The label on the curve for each galaxy is as given in Table \ref{tab:galprops}.  The typical errors of both the dispersion and velocity measurements are 5-15 \kms.  \label{fig:radave}}
\end{figure*}

\begin{figure*}[t]	
\epsscale{1.0}
\plotone{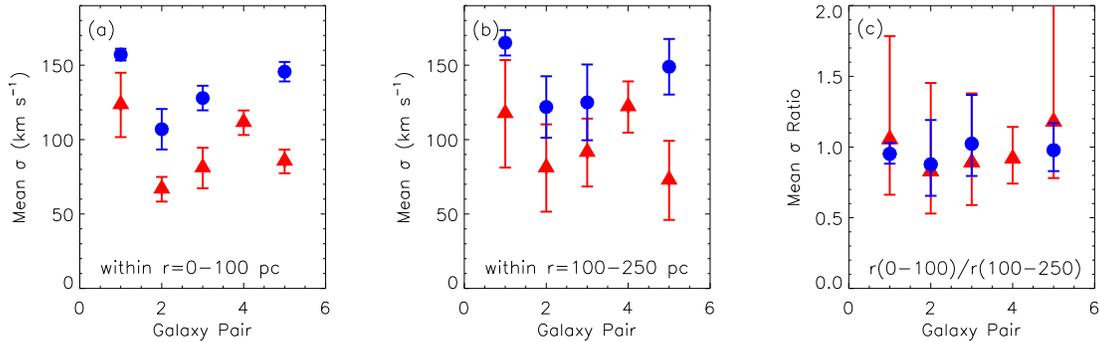}
\caption[]{Mean uniformly weighted CO velocity dispersion of all pixels within apertures of radius (a) 0 to 100 pc and (b) 100 to 250 pc and (c) the ratio of these two.  Triangles represent Seyfert galaxies while circles are quiescent galaxies.  Error bars represent the standard deviation of the dispersion measured in the pixels within the aperture considered.  The probability that the two subsamples are drawn from different parent populations based on both the r=0-100 pc and r=100-250 pc apertures is 93.8\%. \label{fig:pair_cosig}}
\end{figure*}

\begin{figure*}[t]	
\epsscale{1.0}
\plotone{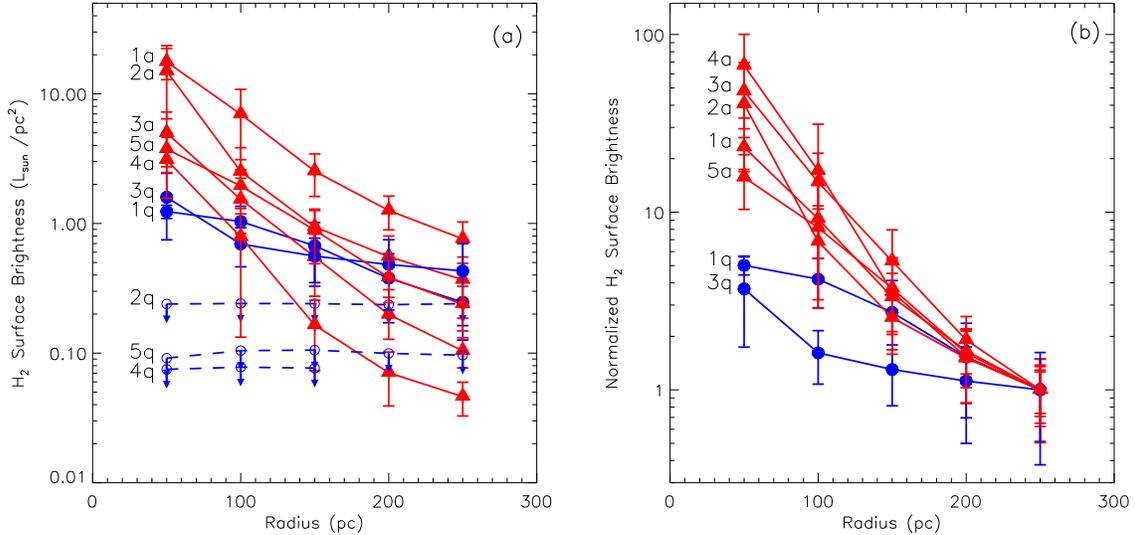}
\caption[]{Mean \htwo\ luminosity as a function of radius (a) intrinsic values and (b) normalized to 250 pc.  Error bars are the standard deviations of measurements within each aperture.  Triangles represent Seyfert galaxies while circles are quiescent galaxies.  Dashed curves and arrows represent upper limits.  The label on the curve for each galaxy is as given in Table \ref{tab:galprops}.  \label{fig:h2_azlum}}
\end{figure*}

\begin{figure*}[t]	
\epsscale{1.0}
\plotone{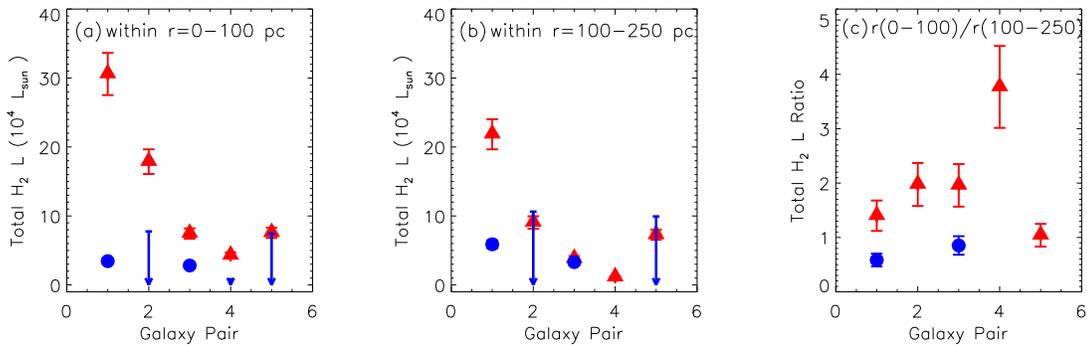}
\caption[]{Total \htwo~luminosity in an apertures of radius (a) 0 to 100 pc and (b) 100 to 250 pc and (c) the ratio of these two.  Triangles represent Seyfert galaxies while circles and upper limits are quiescent galaxies.  Error bars represent the 10\% systematic uncertainty of the flux calibration.  The probability that the two subsamples are drawn from different parent populations based on the r=0-100 pc aperture is 96.9\%. \label{fig:pair_h2lum}}
\end{figure*}

An analysis of the global properties of the matched sample shows some significant differences between the subsamples in properties that are likely related to fueling of BHs in Seyfert galaxies.  All observed properties derived from the line profile analyses described in $\S$ \ref{sec:maps} (flux distribution, velocity, and velocity dispersion) were analyzed in both the stars and molecular gas to identify differences between the Seyfert and quiescent galaxy subsamples.  This was carried out by considering a range of apertures and annuli, as shown in Figures \ref{fig:co_azlum}-\ref{fig:pair_h2lum}.  Circular apertures were used since there is no systematic difference in the subsample large-scale inclinations, and thus the effect of using circular versus elliptical apertures will be similar across the two subsamples.  For the luminosities reported the uncertainties are dominated by the 10\% accuracy of the flux calibration.  For the kinematics (e.g. velocity dispersion), mean values are an average of all bins uniformly weighted (i.e. not luminosity weighted) and uncertainties are taken to be the standard deviation of values within the aperture considered.

For those properties where a difference exists between the Seyfert and matched quiescent subsample, the probability that the two subsamples are drawn from distinct populations is calculated based on a cumulative binomial distribution.  This computes the probability that a given number of galaxies in one subsample will have a measured quantity equal to or greater than the same quantity in the pairs of the other subsample.  For example, if the \htwo~luminosity is found to be higher for Seyferts in at least 3 out of 4 galaxy pairs the probability that these two samples are drawn from different parent population is 68.8\%, and if at least 4 out of 4 are higher in Seyferts the probability is 93.8\%.

Properties for which a significant difference exists between the paired subsamples are discussed below.

\noindent \textbf{Stellar Luminosity $-$} As shown in Figures \ref{fig:co_azlum} and \ref{fig:pair_colum}, beyond a radius of 100 pc all 4 Seyferts have lower luminosities compared to their quiescent galaxy pairs (excluding pair 4 with NGC 628 as discussed in $\S$ \ref{sec:628}) while the luminosities are comparable within a radius of 100 pc.  The average surface brightness (Fig. \ref{fig:co_azlum}) at a radius of 200 pc is found to be about two times lower in Seyferts compared to quiescent galaxies (a ratio of the two subsample means is $<\Sigma_{CO,quiescent}>/<\Sigma_{CO,active}>$ = 1.9 while the mean of the pair-wise ratios is $<\Sigma_{CO,quiescent}/\Sigma_{CO,active}>$ = 2.1).  The total stellar luminosity (Fig. \ref{fig:pair_colum}) within 250 pc is about 1.5 times higher in quiescent galaxies (a ratio of the two subsample means is $<L_{CO,quiescent}>/<L_{CO,active}>$ = 1.5 and the mean of the pair-wise ratios is $<L_{CO,quiescent}/L_{CO,active}>$ = 1.5).  The difference beyond 100 pc shown in Fig. \ref{fig:pair_colum}b implies a 93.8\% probability that the two subsamples are drawn from different populations.  However, the difference in stellar luminosity disappears within a radius of 100pc, where both the total stellar luminosities and average surface brightness are comparable.  Normalizing the surface brightness profiles at a radius of 250 pc (Fig. \ref{fig:co_azlum}b; see also Fig. \ref{fig:pair_colum}c) shows that the stellar luminosity in Seyferts is more centrally concentrated in all Seyferts compared to all quiescent galaxies.  Of note is that the stellar luminosities in the Seyfert subsample have been corrected for AGN contamination based on the dilution of the CO bandhead assuming an intrinsic CO equivalent width of 12\AA~(see $\S$ \ref{sec:agn} and $\S$ \ref{sec:cubefits} for details).  Taking a more extreme intrinsic CO equivalent width of 17\AA, which is the highest CO equivalent width in the \citet{oliva95} sample for a star-forming H II galaxy and lowers the central stellar luminosities by 30\%, the Seyfert stellar luminosity profiles are still found to be more centrally concentrated than those of the quiescent galaxies.

\noindent \textbf{Stellar Dispersion $-$} The stellar velocity dispersion also differs in Seyferts compared to quiescent galaxies, with all 4 pairs included in the analysis having a lower velocity dispersion out to a radius of 200 pc compared to their quiescent counterparts (see Figures \ref{fig:radave}a and \ref{fig:pair_cosig}).  Given these statistics, the probability that the Seyfert galaxies are drawn from a different parent population in which the velocity dispersion is lower compared to the quiescent galaxy population is thus 93.8\%.  The mean stellar dispersion in Seyferts out to a radius of 100 pc is about 1.4 times lower than that in quiescent galaxies ($<\sigma_{CO,quiescent}>/<\sigma_{CO,active}>$ = 1.4 and $<\sigma_{CO,quiescent}/\sigma_{CO,active}>$ = 1.5).  There is an upturn in the velocity dispersion in all but one of the Seyfert galaxies at a radius of about 200-250 pc, which is not seen in the quiescent galaxy population, suggesting that the dispersion in Seyferts is comparable to that in quiescent galaxies at radii shortly beyond this.  

\noindent \textbf{\htwo~Luminosity $-$}  On the scales measured, the \htwo~luminosity is observed to be elevated in all 5 Seyfert galaxies compared to their paired quiescent galaxies out to a radius of at least 250 pc (Figures \ref{fig:h2_azlum} and \ref{fig:pair_h2lum}).  This comparison includes the 3-$\sigma$ upper limits derived for the three galaxies for which no \htwo~detection was observed.  Furthermore, as shown in the normalized surface brightness profiles (Fig. \ref{fig:h2_azlum}b), Seyferts have a steeper luminosity profile compared to the quiescent galaxies.  On average Seyferts have 4 times higher \htwo~luminosity within a radius of 100 pc compared to their quiescent counterparts, with a range in the sample up to 9 times higher luminosity.  The probability that the five Seyferts are drawn from a different parent population in which the \htwo~luminosity is higher compared to the population of quiescent galaxies is 96.9\% (and 93.8\% at radii beyond 150pc where the limited field of view measured for NGC 628 limits the sample to 4 pairs).  

\section{Assessment of Molecular Gas Outflows and the Presence of Ionized Gas}
\label{sec:intspec}

\subsection{No evidence for Bulk Outflows in Molecular Hydrogen}
\label{sec:redshifts}
To determine if there is evidence of bulk outflows in the molecular hydrogen an independent fitting of the redshifts of the \htwo~emission line and the stellar continuum have been performed with the integrated spectra shown in Fig. \ref{fig:intspec}.  Any systematic offset in the derived line and continuum redshifts would imply radial motion of the molecular gas with respect to the nuclear stellar population.  This would reveal any outflows in the molecular gas, such as those often seen in ionized gas on the order of several hundreds of \kms~or more, including in at least some of the Seyferts in this sample (e.g. NGC 6814; \citealt{mueller11}).

The \htwo~emission was fit in the integrated spectrum of each of the 7 galaxies in which it was detected, using a method analogous to that described in $\S$ \ref{sec:cubefits} for fitting of the emission line at each spaxel in the 2D data cubes.  The fit to the stellar continuum was similar to that described for NGC 628 in $\S$ \ref{sec:628}: the rest frame wavelength region of 2.285-2.305 \mic~was fit with a stellar template generated from a stellar library (without a correction for reddening), but in this case assuming a fixed dust fraction and stellar dispersion.  Although this fit is to the entire stellar continuum within the fit wavelength range, the resulting best-fit redshift is dominated by the CO bandheads.  A comparison of the redshifts determined via each of these methods shows that there is no systematic offset in velocity between the molecular gas and stellar continuum, and thus no detected bulk outflow of molecular gas.   Any differences are at the level of about 5 \kms~(except NGC 5643 which is only 18 \kms), and there is no obvious systematic trend in the Seyfert versus quiescent subsamples.  This confirms the conclusions of others that the kinematics of the nuclear molecular gas in Seyfert galaxies is dominated by rotation rather than bulk outflow and is thus a reasonable tracer of the cool ISM (e.g. \citealt{hicks09}, \citealt{sb10}, \citealt{riffelsb11})

\subsection{General Properties of ionized Gas: Br $\gamma$ and Coronal Line Emission}
\label{sec:ionized}

Brief descriptions of the ionized gas detected in the 2D data cubes and integrated spectra shown in Fig. \ref{fig:intspec} are given below, including a comparison of the general properties of this emission in the Seyfert versus quiescent subsamples.  A detailed analysis of the nature of this emission, and its 2D distribution and kinematics, is beyond the scope of this paper, but will be presented in a future paper.

\noindent \textbf{Br $\gamma$ 2.166 \mic~$-$} The presence of Br $\gamma$ emission in a Seyfert galaxy can be interpreted as the result of star formation and/or the AGN activity itself (emission from the narrow and/or broad line regions).  In the absence of other typical signatures of AGN activity (e.g. non-stellar continuum, X-ray emission), the detection of Br $\gamma$ emission is expected to be the result of star formation.  In the observed sample, none of the quiescent galaxies have measurable Br $\gamma$ emission in their integrated spectra or at any location in the 2D data cubes.  This suggests that these galaxies have a relatively old stellar population within the nuclear region.  On the other hand, all five of the Seyfert galaxies have detectable Br $\gamma$ emission.  Furthermore, in each of these cases the 2D distribution of the emission extends beyond the FWHM of the PSF (with a SNR of at least 2, and much higher in several cases), and thus beyond the region significantly influenced by emission from the broad line region.  The emission in the Seyfert galaxies could therefore be due to excitation within the narrow-line region, and thus associated with the AGN activity itself, but a contribution from relatively recent star formation in the nuclear region is also possible.  A future paper will present 2D maps of the Br $\gamma$ emission distribution and kinematics as well as explore the nature of this line emission.

\noindent \textbf{[Ca VIII] $\lambda$ 2.321 \mic~$-$} Also of note is the presence of the AGN coronal line Ca [VIII] emission in at least 4 of the 5 Seyfert galaxies, the exception being Seyfert NGC 7743.  Any detection of this emission in NGC 7743 is marginal, although this is not surprising since, as already noted, there is evidence that this is a weaker AGN than the others in the sample ($\S$ \ref{sec:agn}).  In some cases, this emission is not obvious in the integrated spectra in Fig. \ref{fig:intspec} since it is generally concentrated to the vicinity of the AGN and is thus washed out in the full FOV integrated spectra shown.  In none of the quiescent galaxies is this coronal line emission detected, providing additional confirmation of the lack of AGN activity in these galaxies.

\section{Discussion}
\label{sec:discussion}

We have identified statistically significant differences between Seyfert and quiescent galaxies, and it is these properties unique to Seyfert galaxies that provide the first direct clues to what processes lead to fueling of their central BHs.  We find that Seyferts have a steeper stellar surface brightness profile compared to quiescent galaxies and lower velocity dispersion out to a radius of 200 pc.  In addition, Seyferts have an elevated \htwo~luminosity within this same nuclear region that is more centrally concentrated than the gas distribution seen in quiescent galaxies.  We therefore conclude that we are now probing scales relevant to the fueling of Seyfert galaxies.  \citet{dumas07} also conclude that it is on scales of hundreds to tens of parsecs that a link between host galaxy and BH exists based on their analysis of disk of ionized gas which are likely the counterparts to the cold gas disks detected in \htwo.  This suggests that the mechanisms responsible for the inflow of material are likely to be shaping the observed characteristics of the nuclear (r $\le$ 250 pc) stars and gas.  By understanding how these observed differences arise, combined with future detailed modeling of the observed kinematics, we can begin to form a more complete picture of what processes are dictating the growth of BHs in the local universe and how they relate to coevolution of BH and host galaxy.

\begin{figure*}[t]	
\epsscale{0.5}
\plotone{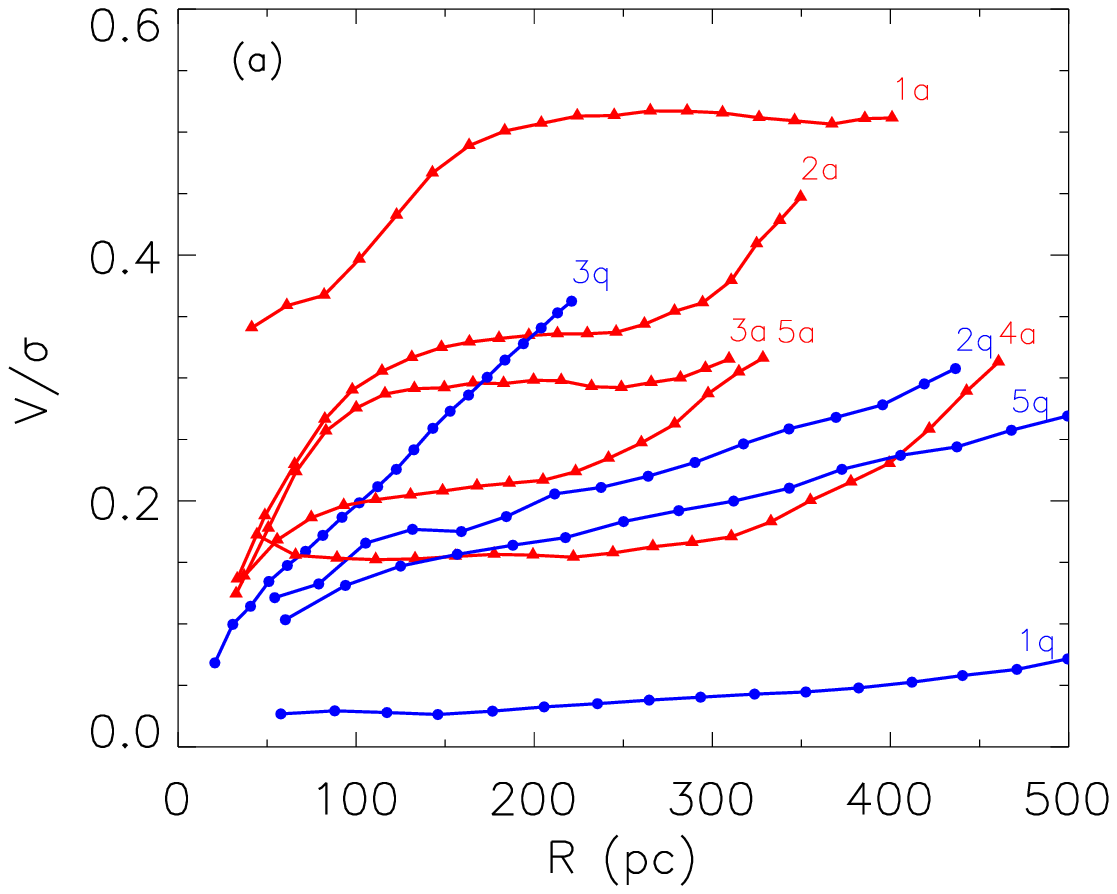}
\plotone{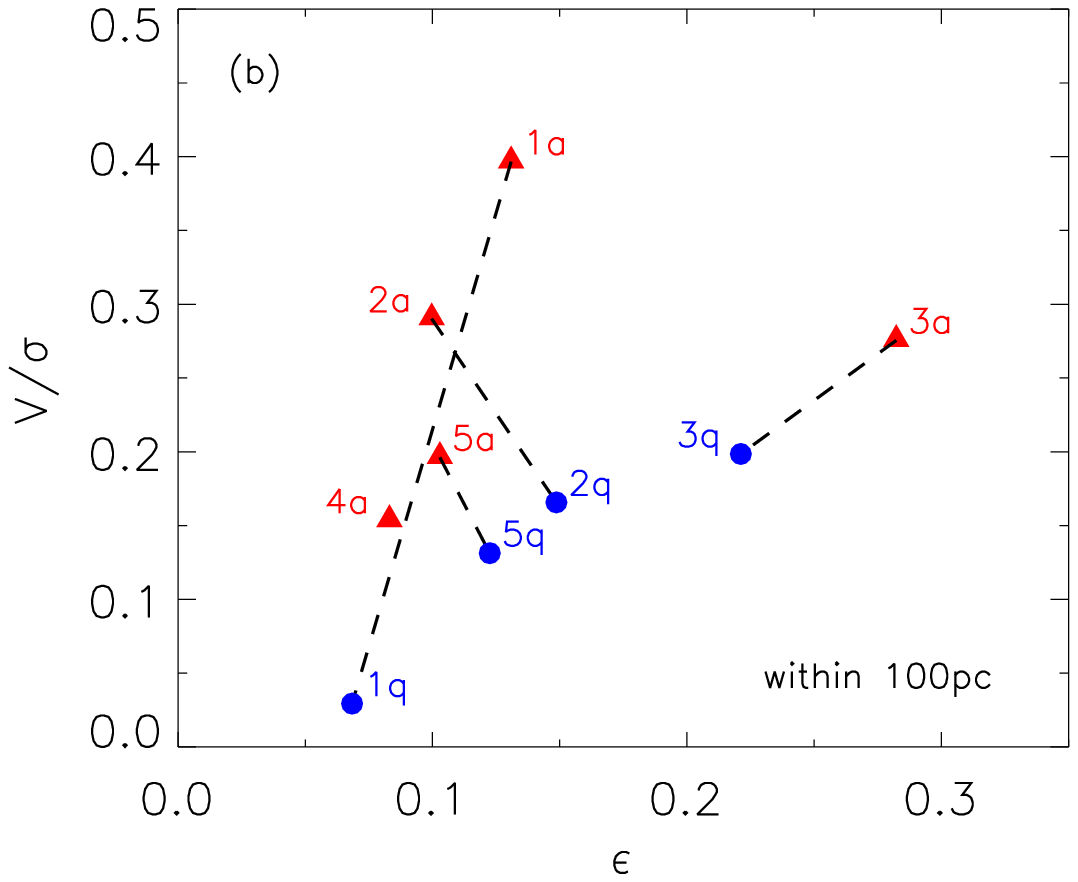}
\caption[]{Global effective CO $V/\sigma$, following \citealt{binney05}, (a) as a function of radius and (b) versus moment ellipticity $\epsilon$ within a radius of 100 pc.  Velocities has been corrected for inclination based on the large scale host galaxy inclination angles reported in \citet{Martini03a}.  Triangles represent Seyfert galaxies while circles are quiescent galaxies.  The label on the curve for each galaxy is as given in Table \ref{tab:galprops}.  \label{fig:global_vs}}
\end{figure*}

\subsection{Significance of Differences in Molecular Gas Properties}
If the \htwo~luminosity reflects gas mass, as it does on large scales (e.g. \citealt{dale05}, \citealt{ms06}, \citealt{hicks09}), then the active galaxies have gas masses on average 4 times higher than similar quiescent galaxies within a radius of 100 pc.  On the other hand, it is known that the \htwo~luminosity on small scales can be strongly influenced by local excitation.  For example, the presence of shocks or a young stellar population can influence the \htwo~emission.  X-ray heating of the gas from the AGN itself is also a possibility which has support both theoretically (\citealt{maloney96}, \citealt{meijerink05}, \citealt{meijerink07}) as well as observationally (\citealt{maloney97}, \citealt{rodriguez05}).  Thus the enhanced \htwo~luminosity could reflect different excitation mechanisms and efficiency (i.e. fraction of hot gas) between the two subsamples.  In either case, the difference in \htwo~luminosity points to a physical difference between active and quiescent galaxies on scales of hundreds of parsecs.

\subsection{Significance of Differences in Stellar Properties}
The lower stellar velocity dispersion in Seyfert galaxies can be understood to be either an indication of a difference in the orbital structure of the nuclear region (e.g. the stars are in a dynamically colder disk-like structure) or a difference in mass distribution (e.g. the nuclear region measured contains less mass). The data presented here indicate that Seyfert and quiescent galaxies  differ in the orbital structure of the nuclear region.  A key piece of evidence leading to this conclusion is that the lower stellar velocity dispersion observed in Seyferts is not accompanied by a lower rotational velocity.  As can be seen in Fig. \ref{fig:radave}b, at all radii measured there is no statistically significant difference in the rotational velocities of the paired galaxies or in a comparison of the subsamples as a whole. Thus a given rotational velocity is accompanied by a lower velocity dispersion in Seyferts than it is in quiescent galaxies, which implies that the contribution from the dynamically hot stellar component is less significant in Seyferts than it is in quiescent galaxies.

\begin{figure*}[t]	
\epsscale{0.8}
\plotone{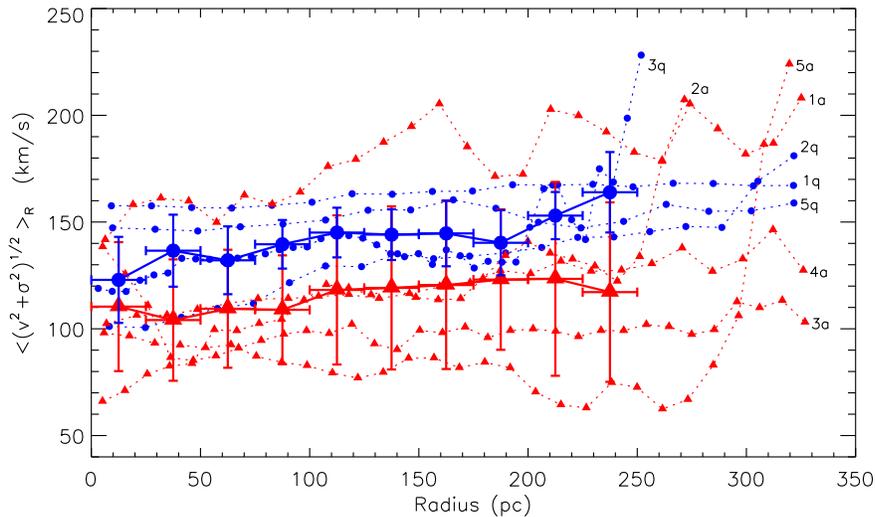}
\caption[]{Mean CO $\sqrt{v^2+\sigma^2}$ as a function of radius.  The CO velocity dispersion is the mean of pixels at each radius bin uniformly weight (rather than luminosity weighted), and the velocity has been corrected for inclination based on the large scale host galaxy inclination angles reported in \citet{Martini03a}.  The larger symbols and solid curve represent the mean of each subsample and the errors bars are the standard deviation of the mean at that radius and the radial bin size.  Triangles represent Seyfert galaxies while circles are quiescent galaxies.  The label on the curve for each galaxy is as given in Table \ref{tab:galprops}.  The typical errors of the $\sqrt{v^2+\sigma^2}$ quantity are 10-20 \kms.  \label{fig:vs_radave}}
\end{figure*}

In an effort to establish the relative contribution of rotational and random velocities in the nuclear region, we consider the effective $V/\sigma$ (a measure of dynamical state of the system in terms of ordered versus random motions; \citealt{binney05}), the specific angular momentum $\lambda_R$ (a measure of dynamical state in terms of the observed projected angular momentum per units mass; \citealt{emsellem07}), and the moment ellipticity $\epsilon$, all of which are generally independent of mass.  Unlike the quantities discussed above, e.g. $\sigma$, which are local values generally within an aperture, these are global values within the stated radius.  All velocities have been corrected for inclination using the large-scale values reported in \citet{Martini03a} (R$_{25}$ in Table \ref{tab:galprops}).  As the radial profiles of the effective $V/\sigma$ shown in Fig. \ref{fig:global_vs}a illustrate, at a radius of 100 pc, each of the four pairs has a higher value for the Seyfert galaxy (the pair including NGC 628 is excluded as justified in $\S$ \ref{sec:628}).  This is also shown in Fig. \ref{fig:global_vs}b where the effective $V/\sigma$ is plotted against $\epsilon$ for the global values within a radius of 100 pc.  The result is the same for the specific angular momentum $\lambda_R$, which is expected given this quantity is related to $V/\sigma$ for regular velocity fields.  At a radius of 200 pc (Fig. \ref{fig:global_vs}a), three of the pairs still have higher values in the Seyferts, and {\em Pair 3} has comparable values (the Seyfert galaxy 3a is just a bit lower than the quiescent galaxy 3q).  Beyond this radius the remaining three pairs for which a comparison can be made have higher values in the Seyferts.  These quantities therefore indicate that, at least within a radius of 200 pc, the subsamples differ in orbital structure, with Seyferts having a dynamically cooler structure compared to the quiescent galaxies.  Outside of a radius of 200 pc the trend for increasing stellar velocity dispersion with radius in the Seyfert galaxies suggests that the subsample differences in orbital structure disappear beyond this radius.

A lower velocity dispersion could also be the result of a systematic subsample difference in inclination, but we consider this unlikely due to (1) a lack of systematic difference in inclination on large scales (see Table \ref{tab:galprops}; mean of the difference in pairs is 7.5$\pm$5.3\deg~lower in Seyferts) and (2) a lack of systematic difference in small-scale inclination derived from fitting of the stellar continua isophots shown in Figures \ref{fig:3227_maps}-\ref{fig:357_maps} (mean of the difference in pairs is 7.8$\pm$5.5\deg~higher in Seyferts).  A lack of systematic difference in the moment ellipticities (Fig. \ref{fig:global_vs}b) provides additional support for the conclusion that it is not an issue of inclination that results in the observed differences in velocity dispersion.  In addition, the subsample differences in effective $V/\sigma$ and specific angular momentum $\lambda_R$ cannot be explained via a systematic difference in inclination since there is no systematic difference in the observed moment ellipticity $\epsilon$.  It is therefore more likely that a difference in orbital structure explains the observed differences in $V/\sigma$ and $\lambda_R$ rather than a systematic difference in inclination. 

In order to assess the mass in the nuclear region, the quantity $\sqrt{v^2+\sigma^2}$ is considered (Fig. \ref{fig:vs_radave}).  This quantity gathers the first velocity moments so that a decrease in the dispersion should be partly compensated by an increase in velocity if the difference in dispersion were fully due to a variation in the enclosed mass (\citealt{binney08}).  As before, all velocities have been corrected for inclination using the large-scale values reported in \citet{Martini03a} (R$_{25}$ in Table \ref{tab:galprops}).  Throughout the region measured, the difference between the Seyfert-quiescent pairs is not statistically significant when considering $\sqrt{v^2+\sigma^2}$.  The only pair with a significant difference in this quantity is {\em Pair 5} in which the quiescent galaxy has a higher value throughout the majority of the FOV measured.  In the remaining pairs the difference between the Seyfert and quiescent galaxies is within the error of the measurement (a difference of less than 25 \kms).  There is also no significant difference in the subsample means (solid curves in Fig. \ref{fig:vs_radave}).  This lack of subsample difference in $\sqrt{v^2+\sigma^2}$ supports the theory that the lower stellar velocity dispersion observed in Seyferts galaxies is not driven by a difference in mass enclosed within the nuclear region.

In addition to kinematical considerations, a number of photometric quantities suggest that the lower stellar velocity dispersion in the Seyfert subsample can be at most only partially explained by a systematic difference in the subsample mass within the nuclear region.  If the observed subsample difference in dispersion of a factor of 1.4 within a radius of 100 pc is due to a difference in mass, and if it is assumed that mass is a function of $\sigma^2$, then a mass difference of a factor of two is required.  The difference implied by the lower total stellar luminosity in Seyferts within a radius of 250 pc (Fig. \ref{fig:pair_colum}; see also discussion in $\S$\ref{sec:global}), if taken to reflect mass and assuming a fixed mass-to-light ratio, is a mass difference of a factor of only 1.5$\pm$0.5.  Considering the {\em H}-band 2MASS estimated total photometry (Table \ref{tab:galprops2}) as an indicator of stellar mass, the mean difference in the subsamples is 0.15 $\pm$ 0.37 mag, with Seyferts slightly fainter (and thus less massive).  This suggests a mass difference of a factor of 1.15$^{+0.46}_{-0.33}$, also less than the full difference of 2 required to explain the lower stellar dispersion.  A comparison of the subsample galaxy sizes, quantified in this case as half of D$_{25}$ (Table \ref{tab:galprops2}), reveals a subsample difference with Seyferts being 23$\pm$13\% smaller, and thus less massive, than the quiescent galaxies.  However, a systematic difference in D$_{25}$ or {\em H}-band photometry does not necessarily reflect a difference in actual nuclear mass since these quantities are significantly influenced, if not dominated, by properties of the stellar disk.  A comparison of the bulge-to-total flux ratios estimated based on the Hubble type (see Table \ref{tab:galprops2} for details) indicates there is no statistically significant difference between the subsamples.  Therefore the difference in disk mass  implied by subsample differences in D$_{25}$ and {\em H}-band photometry, as well as the difference indicated by the total stellar luminosities, suggest that the total nuclear masses may differ, but not significantly enough to explain the observed difference in velocity dispersion.  In addition, the steeper gradient of the central stellar surface brightness observed in the Seyfert galaxies compared to the quiescent galaxies cannot be easily explained through a difference in mass and suggests a more complex situation.    

Having ruled out a systematic difference in inclination and finding minimal evidence for a subsample difference in mass within the nuclear region, we conclude that the observed subsample difference in stellar velocity dispersion is due primarily to a difference in orbital structure.  In this scenario Seyfert galaxies have a dynamically cooler system within a radius of 200 pc in comparison to quiescent galaxies which results in the lower observed nuclear stellar velocity dispersions.

\subsection{Evidence for ISM Inflow}

The ways in which Seyfert galaxies are observed to differ from quiescent galaxies, which presumably also harbor central BHs that are in a dormant state, suggests that the fueling of an AGN is accompanied (or triggered) by an altered nuclear region that extends out to a radius of 200-250 pc.  The observed lower stellar velocity dispersion within this region appears to be due to a comparatively dynamically cooler disk-like orbital structure in Seyfert galaxies.  These so called ``$\sigma$-drops'', can be understood as evidence for the presence of an additional nuclear stellar component in Seyfert galaxies that is dynamically cold in comparison to the (dynamically hotter) stars that dominate the detected stellar light in the quiescent galaxies (\citealt{emsellem01},\citealt{wozniak03},\citealt{wozniak06}).  Similar ``$\sigma$-drops'' have been observed in a number of AGN galaxies (e.g. \citealt{emsellem08}, \citealt{comeron08}, \citealt{riffel09,riffel10,riffel11}), and there is also evidence that a central decrease in stellar dispersion is correlated with Seyfert nuclear activity (\citealt{dumas07}, \citealt{comeron08}).  This interpretation of the lower stellar velocity dispersion as a dynamically cold nuclear component is consistent with the more centrally concentrated stellar distribution seen in the Seyfert stellar light profiles: as the dynamically cold nuclear component begins to dominate the light within the central 200 pc the profile represents that of the nuclear component rather than the higher velocity dispersion stars and thus steepens.  In quiescent galaxies such a nuclear stellar structure may either be absent or be much harder to detect because its stellar population no longer dominates the near-IR light over the higher velocity dispersion component (e.g. the dynamically cold stars are older and fainter in comparison to that found in Seyferts).  

A consequence of interpreting the observed differences in subsample stellar properties as evidence for a nuclear dynamically cold component in Seyferts is that a reservoir of gas must be present from which the stellar population of this component can form, and thus inflow of ISM must occur to supply the gas to the nuclear region.  The presence of such a nuclear component thus requires that ISM inflow has occurred.  In this scenario the observed enhanced \htwo~emission and greater central concentration in Seyferts can then naturally be interpreted as evidence of an enhanced nuclear gas mass rather than being explained solely by a difference in excitation or efficiency.  The central BH can then be fueling either directly from this supply of nuclear gas or indirectly through smaller scale processes, including from stellar winds (\citealt{Davies07}, \citealt{schartmann10}).  In such a scenario, the Br $\gamma$ emission, which was detected in only the Seyfert subsample, may be partially due to excitation from young/hot stars within this nuclear region of enhanced gas mass and nuclear star formation. 

Also of note is that in galaxies from both subsamples there is evidence of non-circular velocities (e.g. radial flow) in the velocity fields of \htwo, despite rotation generally dominating the gas motion.  This can be easily seen in Fig. \ref{fig:3227_maps}-\ref{fig:357_maps} through a comparison of the stellar zero iso-velocity contours with that of \htwo, which is often twisted in the nuclear region as expected in the presence of a nuclear bar or spiral.  Detailed modeling and characterization of the nuclear kinematics of each galaxy will be presented in a future paper and any additional subsample differences will be identified to determine what fueling mechanisms contribute to the fueling of the AGN present in the Seyfert subsample.

\section{Conclusions}
\label{sec:conclusions}
With a sample of matched Seyfert-quiescent galaxy pairs we have started to identify the processes that drive BH growth in the local universe, as well as the processes potentially at work at higher redshifts, by directly measuring the morphology and kinematics of the stars as well as the warm molecular gas that traces the ISM making up the reservoir of AGN fuel.  We find subsample differences within the central kpc that are correlated with AGN activity.  The subsample differences indicate that Seyferts have (1) a more centrally concentrated nuclear stellar surface brightness with a lower stellar luminosity outside of a radius of 100 pc, (2) a lower stellar velocity dispersion within a radius of 200 pc, (3) elevated \htwo~luminosity out to a radius of at least 250 pc, and (4) more centrally concentrated \htwo~surface brightness profiles.  Although this sample is rather small, comprising of 5 Seyfert-quiescent galaxy pairs, the subsample differences are statistically significant.  This suggests that with these observations of the central 1 kpc down to 50 pc we are now probing scales relevant to the link between the host galaxy and BH growth.  In addition, Br $\gamma$ emission is detected in all 5 Seyfert galaxies and the coronal line [Ca VIII] is present in 4 of the 5 Seyferts.  In contrast, both emission lines are undetectable in the spectra of all of the quiescent galaxies, confirming the lack of AGN activity in these galaxies.

A likely interpretation of the observed subsample differences is that in Seyfert galaxies a nuclear component composed of both stars and gas is present that is either undetected or not present in quiescent galaxies.  This is possible if the nuclear component dominates the detected stellar light out to a radius of 100-250 pc and is dynamically cold in comparison to stars detected in quiescent galaxies.  The elevated \htwo~luminosities and higher central concentration would then represent a reservoir of gas mass available for forming the nuclear stellar component and potentially fueling the BH.  Assuming the enhanced \htwo~is a reflection of gas mass (as opposed to being entirely due to a difference in excitation mechanisms), this then implies a factor of 4 difference on average in gas mass between Seyfert and quiescent galaxies within a radius of 100 pc.  Under this scenario the nuclear disk-like structure observed is generated by inflow to the nuclear region and likely fueling the BH either directly or indirectly through smaller-scale processes occurring within this nuclear disk.  Another possible explanation of the observed subsample differences is that Seyferts have less mass within a radius of 250 pc compared to quiescent galaxies, however we find minimal evidence for this scenario.

Future work will include a detailed analysis of the kinematics observed in the sample and a subsample comparison will further identify any additional properties unique to Seyfert galaxies.  Modeling of the observed kinematics will then classify (e.g. identify nuclear spirals, bars) and quantify any on-going inflow of gas within the central kpc.  With an understanding of how Seyferts differ in comparison to quiescent galaxies in their nuclear stellar and gas characteristics, and most critically in their inflow of gas, on scales from 500 down to 50 pc we can begin to identify what processes are essential to BH growth in local galaxies and then explore how this might extend to higher redshift systems and impact the general evolution of the host galaxy. 

\acknowledgments
E. K. S. H. acknowledges support from the National Science Foundation Astronomy and Astrophysics Postdoctoral Fellowship under award AST-1002995, as well as support from the NSF Astronomy and Astrophysics Research Grant under award AST-1008042.

\end{document}